\documentclass[useAMS,usenatbib,usegraphicx,a4paper]{mn2e}
\usepackage{aas_macros}
\usepackage{graphicx}
\newcommand{\um}{\ensuremath{\,\mathrm{\mu m}}}
\newcommand{\myr}{\ensuremath{\,\mathrm{Myr}}}
\newcommand{\gyr}{\ensuremath{\,\mathrm{Gyr}}}
\newcommand{\msun}{\ensuremath{\,\mathrm{M_{\odot}}}}

\newcommand{\zsun}{\ensuremath{\,\mathrm{Z_{\odot}}}}
\newcommand{\sfr}{\ensuremath{\,\mathrm{M_{\odot}\,yr^{-1}}}}
\newcommand{\sfrarea}{\ensuremath{\,\mathrm{M_{\odot}\,yr^{-1}\,kpc^{-2}}}}
\newcommand{\zth}{z\ensuremath{\,\sim\,}3}
\newcommand{\zfv}{z\ensuremath{\,\sim\,}5}
\newcommand{\zopen}{z\ensuremath{\,\sim\,}}

\title[Lyman-break galaxies at \zfv]{Lyman-break galaxies at \zfv\ -
  I. First significant stellar mass assembly in galaxies that are not
  simply \zth\ LBGs at higher redshift} 
\author[Verma, Lehnert,
  F\"orster Schreiber, Bremer and Douglas]{Aprajita
  Verma$^{1,2}$\thanks{averma@astro.ox.ac.uk (AV);
    matthew.lehnert@obspm.fr Present affiliation: Laboratoire Galaxies
    Etoiles Physique et Instrumentation, Observatoire de Paris, 5
    Place Jules Janssen, 92195 Meudon, France (MDL);
    forster@mpe.mpg.de (NMFS); m.bremer@bristol.ac.uk (MNB);
    laura@star.bris.ac.uk (LD)}, Matthew D. Lehnert$^{1}$, Natascha
  M. F\"orster Schreiber$^{1}$, 
\newauthor Malcolm N. Bremer$^{3}$,
  Laura Douglas$^{3}$\\ 
$^{1}$ Max Planck Institut f\"ur
  extraterrestrische Physik, Postfach 1612, D-85741 Garching, Germany
  \\ 
$^{2}$ Oxford Astrophysics, Department of Physics, University of
  Oxford, Denys Wilkinson Building, Keble Road, Oxford OX1 3RH,
  UK\\ 
$^{3}$ H H Wills Physics Laboratory, University of Bristol,
  Tyndall Avenue, Bristol BS8 1TL, UK}

\begin{document}

\date{Accepted 2006 December 29. Received 2006 December 07; in original form 2006
October 03}

\pagerange{\pageref{firstpage}--\pageref{lastpage}} \pubyear{2007}

\maketitle

\label{firstpage}

\begin{abstract}

  We determine the ensemble properties of \zfv\ Lyman break galaxies
  (LBGs) selected as V-band dropouts to $i_{AB}<$\,26.3 in the Chandra
  Deep Field South using their rest-frame UV-to-visible spectral
  energy distributions. By matching the selection and performing the
  same analysis that has been used for \zth\ samples, we show clear
  differences in the ensemble properties of two samples of LBGs which
  are separated by 1\gyr\ in lookback time. We find that \zfv\ LBGs
  are typically much younger ($<$100\myr) and have lower stellar
  masses ($\sim$10$^9$\msun) than their \zth\ counterparts (which are
  typically $\sim$\,few$\times$10$^{10}$\msun\ and
  $\sim$320\myr\ old).  The difference in mass is significant even
  when considering the presence of an older, underlying population in
  both samples.  Such young and moderately massive systems dominate
  the luminous \zfv\ LBG population ($\ga$\,70 per cent), whereas they
  comprise $\la$\,30 per cent of LBG samples at \zth. This result,
  which we demonstrate is robust under all reasonable modelling
  assumptions, shows a clear change in the properties of the luminous
  LBGs between \zfv\ and \zth. These young and moderately massive
  \zfv\ LBGs appear to be experiencing their first (few) generations
  of large-scale star formation and are accumulating their first
  significant stellar mass. Their dominance in luminous LBG samples
  suggests that \zfv\ witnesses a period of wide-spread, recent galaxy
  formation.  As such, \zfv\ LBGs are the likely progenitors of the
  spheroidal components of present-day massive galaxies. This is
  supported by their high stellar mass surface densities, and is
  consistent with their core phase-space densities, as
  well as the ages of stars in the bulge of our Galaxy and other
  massive systems.  With implied formation redshifts of
  z$\,\sim\,$6~-~7, these luminous \zfv\ LBGs could have only
  contributed to the UV photon budget at the end of
  reionisation. However, their high star formation rates per unit area
  suggest these systems host outflows or winds that enrich the intra-
  and inter-galactic media with metals, as has been established for
  \zth\ LBGs. Their estimated young ages are consistent with
  inefficient metal-mixing on galaxy-wide scales.  Therefore these
  galaxies may contain a significant fraction of metal-free stars as
  has been previously proposed for \zth\ LBGs
  \citep{2006Natur.440..501Jimenez}.

\end{abstract}


\section{Introduction}

One of the fundamental open questions in cosmology is when did
galaxies form their first generations of stars? Identifying and
studying such galaxies are key steps towards understanding the
physical processes that drive galaxy formation. Probing the formation
and early growth of systems similar to the Milky Way requires
observations of galaxies when the Universe is still young.
Specifically, theoretical calculations \citep{2002MNRAS.336..112Mo}
predict that the most rapid growth of galaxies with masses comparable
to that of the Milky Way occurs at redshifts of approximately
\zfv. This is not long after the end of the (complex) reionisation
process, which recent three-year WMAP results indicate was underway at
z$\,\sim\,$11 \citep{2006astro.ph..4447Alvarez} and was largely
complete by redshifts z$\,\sim\,$6-6.5
\citep{2001AJ....122.2833Fan,2001AJ....122.2850Becker,2004ApJ...617L...5Malhotra,2005astro.ph.12082Fan}.
Estimates of the ionising photon density in the early Universe
suggests that the UV emission from currently known high redshift
galaxies, whether star-formation or AGN dominated, is insufficient to
have caused reionisation. However, these galaxies must have had an
impact on the intergalactic medium at high redshift. Only through
comprehensive studies of the physical properties (including masses,
star formation rates and histories, and clustering strength) of
galaxies that were in place at this epoch, can we accurately assess
their contribution to the mass growth of galaxies like our own, and
their effect on the gaseous intergalactic medium (IGM) at the end of
reionisation.

As part of an ongoing study of high redshift galaxies, we have
investigated the rest-frame UV-optical properties of candidate LBGs at
4.6\,$\la$\,z\,$\la$\,6 selected as V-dropouts using the now standard
Lyman-break technique
\citep{1993AJ....105.2017Steidel,1995AJ....110.2519Steidel,1999ApJ...519....1Steidel}.
Through spectroscopic confirmation, we have successfully demonstrated
the efficacy of this technique for unambiguously identifying
\zfv\ galaxies from deep imaging surveys in the rest-frame UV using
8m-class telescopes \citep{2003ApJ...593..630Lehnert}. From these data
it has been possible to determine the comoving density
\citep{2004MNRAS.355..374Bunker}, the likely contribution to the end
of reionisation \citep{2003ApJ...593..630Lehnert}, and the fraction of
sources which host super-massive black holes
\citep{2004MNRAS.347L...7Bremer} of \zfv\ LBGs. However, the
rest-frame UV data alone are thus far insufficient to accurately
constrain the ages, dust content, star formation rates and masses of
the LBG population at high redshift.

Rather, accurately constraining these parameters relies on well
measured rest-frame UV-to-optical SEDs.  At \zfv, the rest-frame UV
emission from ongoing star-formation is redshifted into the observed
visible, while emission in the rest-frame visible to near-infrared
from older stars, diagnostic of longer or earlier periods of star
formation, shifts into the mid-infrared.  Several fields now have
excellent multi-wavelength data sets from both the Hubble and Spitzer
Space Telescopes (HST and Spitzer, respectively), supplemented by
ground-based data sets, forming an ideal basis for selecting and
studying samples of distant galaxies across the full rest-frame UV to
visible wavelength range.

The most detailed multiwavelength studies to date of 5\,$<$\,z\,$<$\,7
LBGs have been centered upon a small fraction of dropouts that benefit
from amplification due to lensing and/or those detected with
Spitzer-IRAC
\citep{2005ApJ...618L...5Egami,2005MNRAS.362.1054Schaerer,2005ApJ...634..109Yan,2005ApJ...635..832Mobasher,2005ApJ...635L...5Chary,2005MNRAS.364..443Eyles,2005ApJ...630L.137Dow,2006astro.ph..4554Yan}.
Since, at z\,$>$\,5, the Balmer-break lies between the $K_s$ and the
IRAC pass-bands, the IRAC data are highly effective in constraining the
ages of these systems. A strong detection with IRAC normally confirms
the presence of a Balmer break or an intrinsically more UV luminous
system.  The derived properties of these high redshift LBGs are
fascinating; massive (few-several $\times$\,10$^{10}$\msun) and
strongly star forming systems with ages comparable to the age of the
Universe at that epoch.  The presence of such massive and evolved
systems at high redshift, in place less than a billion years after the
Big Bang, challenges the expectations from bottom-up hierarchical
structure formation scenarios \citep[however, also
  see][]{2006MNRAS.tmp..944McLure}.

These derived masses and ages of z\,$\ga$\,5 LBGs are similar to the
average properties of LBGs at redshift 3.  Following the pioneering
work of Steidel et al., the last decade has seen intensive
observational and theoretical studies on the properties of LBGs at
\zth, providing a wealth of information about this abundant population
of UV-bright galaxies at this epoch.  These star-formation dominated
systems are seen to host strong outflows
\citep{2001ApJ...554..981Pettini,2003ApJ...584...45Adelberger,2003ApJ...588...65Shapley},
typically have sub-solar-to-solar metallicities
\citep[$\sim$0.3\,-\,1\,Z$_{\odot}$;][]{2001ApJ...554..981Pettini,2003ApJ...588...65Shapley}
and possibly contain a significant component of metal-free (population
III) stars \citep{2006Natur.440..501Jimenez}. LBGs are shown to be
highly clustered on both large and small scales, the latter being
indicative of common halo objects
\citep{1998ApJ...503..543Giavalisco,1998ApJ...505...18Adelberger,2004ApJ...611..685Ouchi,2005ApJ...635L.117Ouchi,2006ApJ...642...63Lee}.
Recent follow-up of \zth\ LBGs with Spitzer has revealed a division in
the population between LBGs with significant dust-attenuated
star-forming regions as well as the UV-emitting ones
\citep{2005ApJ...634..137Huang}.  This division is also seen for LBGs
at z$\,\sim\,$1 \citep{2006A&A...450...69Burgarella} with 40 per cent being
infrared bright.  Recently \citet{2005ApJ...633..748Reddy} have
reported on the potential relation between UV- and IR selected
populations.  \citet{2005ApJ...634..137Huang} suggest IR-luminous LBGs
are the missing link between LBGs and sub-mm galaxies and are the
progenitors of present-day giant ellipticals \citep[see
  also][]{2000ApJ...544..218Adelberger,2001MNRAS.327..895Shu,2004ApJ...611..685Ouchi,2006astro.ph..5355Rigopoulou}.

But how do these properties compare to those samples of LBGs at higher
redshifts? Do we see an evolution in the properties of similar LBGs at
these two epochs?  Because of the small number of sources
investigated, and the inhomogeneity of the selection criteria used, it
has not yet been possible to consistently compare the derived
properties of z\,$\ga$\,5 LBGs directly with those of Lyman-break
galaxies at \zth. In a preliminary study,
\citet{2004ApJ...610..635Ando} find that unlike \zth\ LBGs
\citep{2003ApJ...588...65Shapley} the majority of luminous \zfv\,-\,6
LBGs have weak or absent Lyman alpha (Ly$\alpha$) emission and strong
low-ionisation absorption lines, already indicating key differences
between the populations at redshift 5 and redshift 3. We complement
this comparison by analysing the properties of a large and reliable
sample of the most luminous LBGs (L$\,>\,$L*) at \zfv\ which match the
luminosity to which LBGs at redshift 3 have been spectroscopically
confirmed.  This matched selection permits a direct comparison between
the properties of similarly luminous LBGs in place in a
$\sim$1.2\gyr\ old Universe and those at redshift 3,
$\sim$1\gyr\ later. This difference is longer than the typical
duration of the UV luminous phase of a \zth\ LBG, thus we are not
comparing the same galaxies, but galaxies at two epochs with similar
UV-emission characteristics.

To address these issues, we present the results from an analysis of
the properties of a sample of luminous V-band dropouts selected from
the multiwavelength datasets of the Chandra Deep Field South
\citep[CDFS,][]{2001ApJ...551..624Giacconi}. This field was chosen as
the southern field of the Great Observatories Origins Deep Survey
\citep[GOODS,][]{2003mglh.conf..324Dickinson}, for which public
imaging data are available over ten bands covering the optical
(HST-ACS), near-infrared (VLT-ISAAC) and mid-infrared (Spitzer-IRAC)
wavelength ranges. In section \ref{sample} we describe our methodology
and selection criteria that define this robust sample of
\zfv\ LBGs. The stellar evolutionary modelling is discussed in section
\ref{model}. The resultant physical properties of mass, ages, star
formation rate and extinction of the galaxies in this sample are
detailed in Section \ref{results}\footnote{We explore how different
  assumptions in the modelling process affects the results of our
  analysis in Appendix A and the potential contribution from an
  underlying old population in Appendix B.} and the comparison to the
properties of \zth\ galaxies is presented in Section
\ref{comp}. Unlike similarly luminous \zth\ LBGs, and other z$\,>\,$5
LBG samples, we find our sample to be dominated by young ($<$100\myr)
and moderately massive ($\sim 10^{9}$\msun) galaxies. We discuss the
implications of our findings in Section \ref{discussion}.

We adopt the following flat cosmology throughout this paper:
$H_{0}$=70kms$^{-1}$Mpc$^{-1}$, $\Omega_{m}$=0.3,
$\Omega_{\Lambda}$=0.7. All magnitudes are based on the AB magnitude
system \citep{1983ApJ...266..713Oke}.

\section{Data and Sample Selection}
\label{sample}

\subsection{Selection Criteria}
We applied a Lyman-break colour selection to the publicly available
HST/Advanced Camera for Surveys (ACS) imaging datasets of the CDFS
obtained through the filters F435W (\textit{B}), F606W (\textit{V}),
F775W (\textit{i}) and F850LP (\textit{z})
\citep{2004ApJ...600L..93Giavalisco}. Specifically, we selected
objects with $V_{AB}$-$i_{AB}>1.7$\,mag which were not detected
(signal-to-noise ratio $<$ 3) in the F435W ($B$-band) image (\rm{i.e.}
short-ward of the 912\AA\ Lyman-break at \zfv) ensuring that we only
selected sources with a clear Lyman discontinuity in their
emission. We then required that the objects were reliably detected
(signal-to-noise ratio $>$ 5 in the $z$-band) in order to accurately
measure the sizes of the UV emitting regions.  Furthermore, as
mentioned above, we matched our sample to the magnitude limit to which
LBGs at \zth\ have been spectroscopically confirmed.  This limit of
$\cal{R}_{AB}$$\sim$25.5 probes \zth\ LBGs brighter than
$\sim\,0.4\,$L*$_{z=3}$ \citep{2003ApJ...592..728Steidel}. At redshift
\zfv\, $\cal{R}_{AB}$$\sim$25.5 corresponds to $i_{AB}$$\sim$26.3\,mag,
the magnitude to which we select our sample of \zfv\ LBGs. This colour
selection is similar to those previously used to select $V$-band dropouts
(or the analogous $R$-band dropouts in \citealt{2003ApJ...593..630Lehnert},
but see also \citealt{2006astro.ph..1367Vanzella, 2006astro.ph..4250Stark}
for a discussion of alternative colour cuts), which have been
successfully spectroscopically confirmed to lie at \zfv\
\citep{2004ApJ...600L.103Giavalisco,2005ApJ...634..109Yan,
2006astro.ph..1367Vanzella}.

\subsection{Input data}

\begin{figure*}
\begin{centering}
\begin{tabular}{cc}
  \multicolumn{2}{c}{\includegraphics[width=0.9\textwidth]{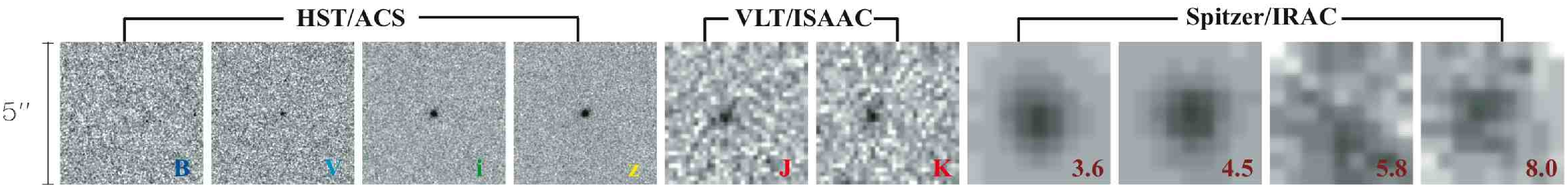}}\\
  \includegraphics[width=0.45\textwidth]{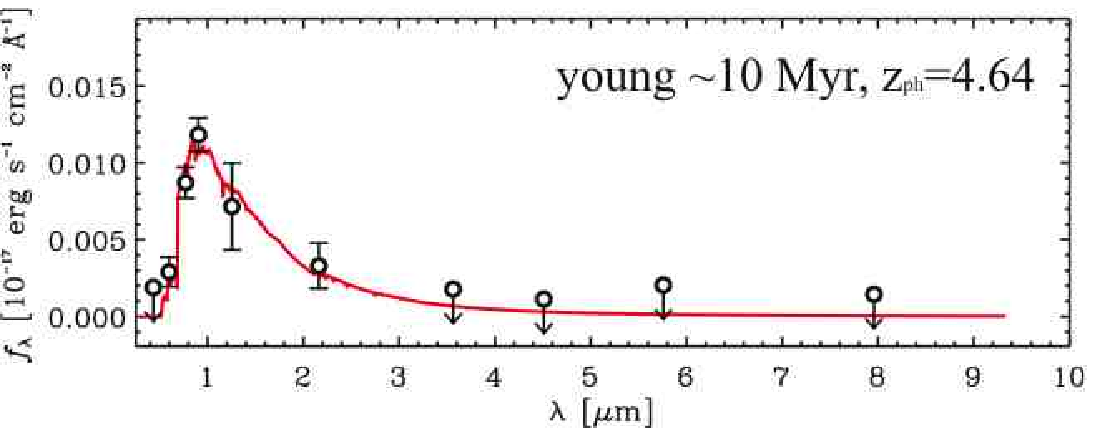} & \includegraphics[width=0.45\textwidth]{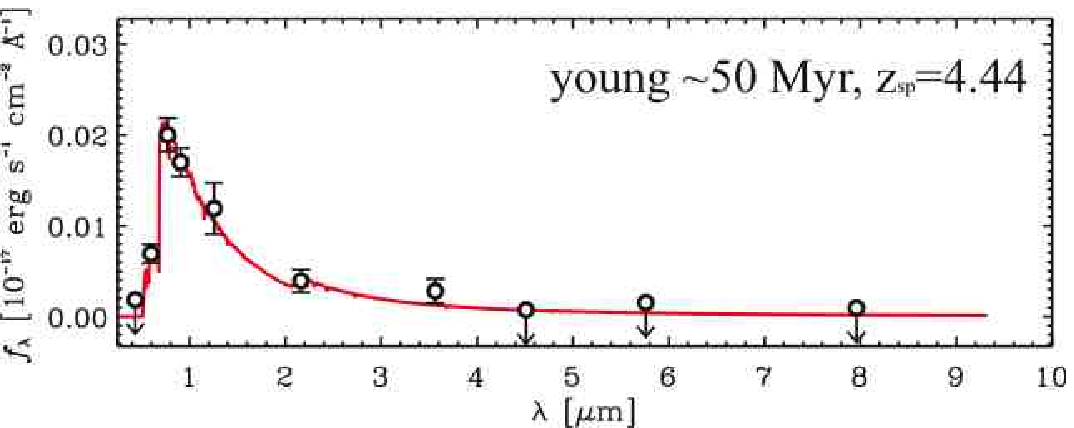}\\
  \includegraphics[width=0.45\textwidth]{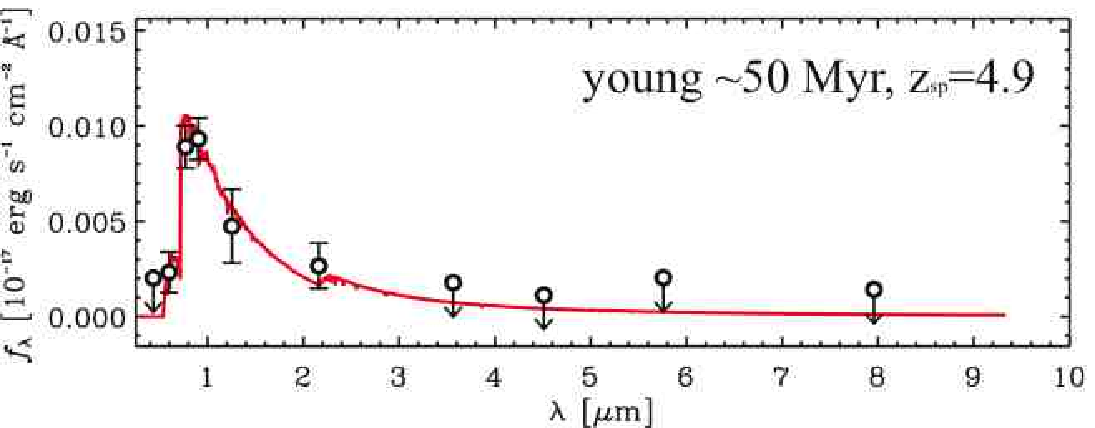} & \includegraphics[width=0.45\textwidth]{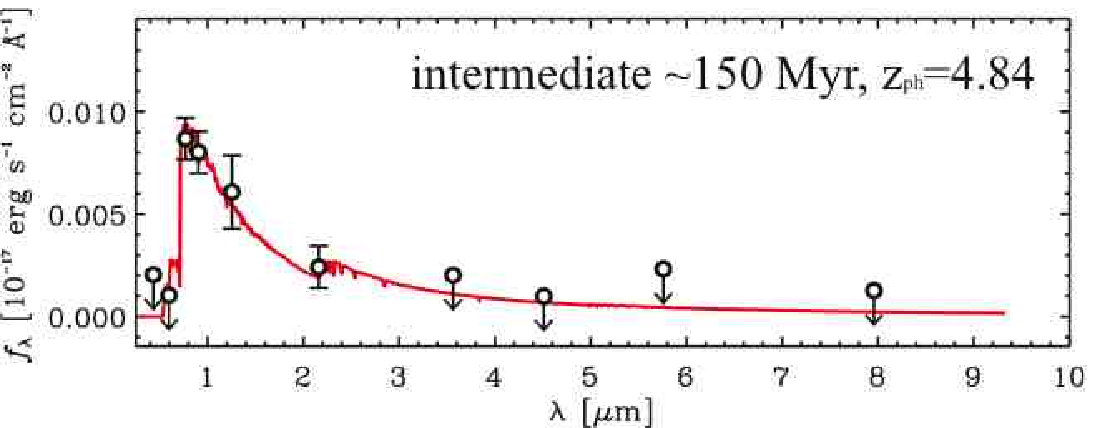}\\
  \includegraphics[width=0.45\textwidth]{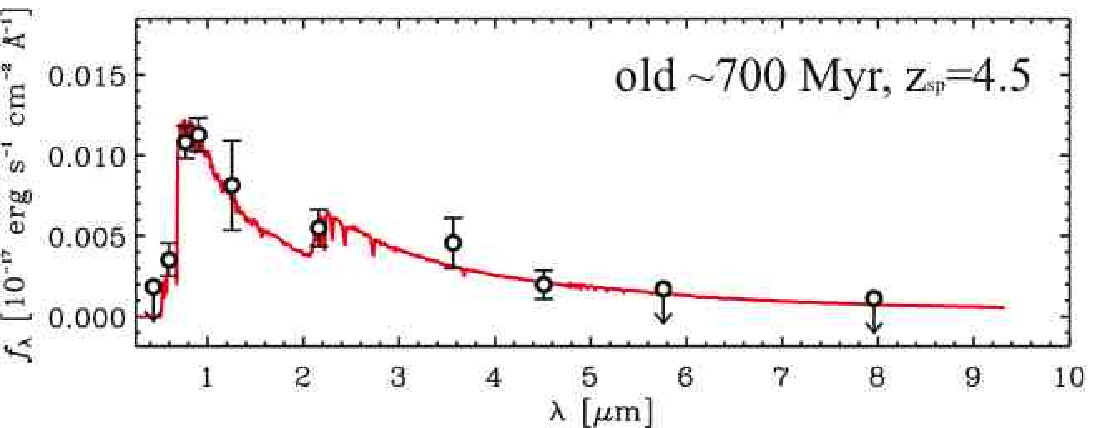} & \includegraphics[width=0.45\textwidth]{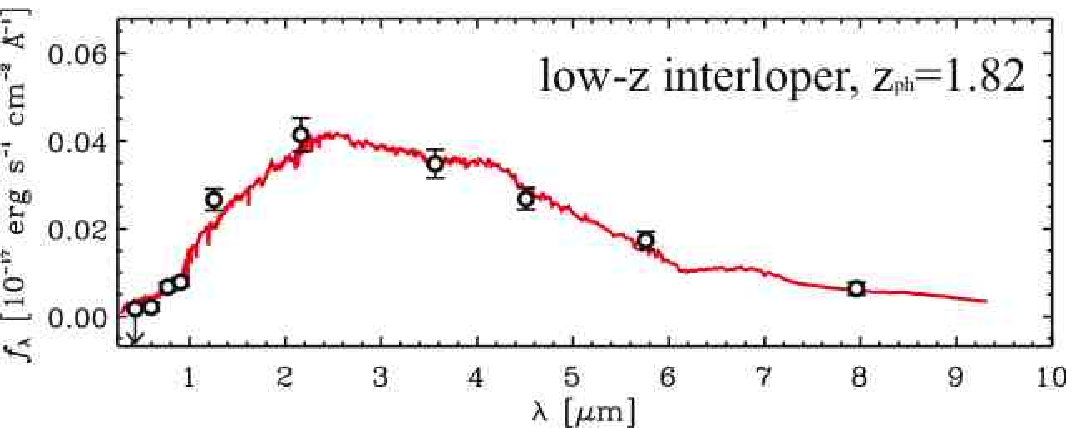}\\
\end{tabular}
\caption{(i) This figure shows 5"$\times$5" images of a typical
    galaxy from our sample in the 10 filter-bands we consider. 
    Six
    spectral energy distributions (SEDs) are shown in panels
    (ii-vii). The best-fit model is plotted over the data
    points. Panels (ii-vi) show SEDs for five galaxies from our robust
    sample of \zfv\ LBGs. The SEDs are arranged (from left to right
    then down) in approximate best-fit age order showing the emergence
    of the Balmer-break between the $K_s$ and 3.6$\mu$m data
    points. The SED of a low redshift interloper is shown in panel vii
    for comparison. Panels iii, iv and vi are spectroscopically
    confirmed \zfv\ LBGs.}
\label{fig:seds}
\end{centering}
\end{figure*}

For each object we combined the ACS data with publicly available near-
and mid-infrared datasets to construct multi-wavelength SEDs covering
0.45 to 8$\mu$m (Figure \ref{fig:seds}). The ground-based $J$- and
$K_{s}$-band data were taken with ISAAC at the Very Large Telescope
(VLT) \citep[][and
  http://www.eso.org/science/goods/imaging/products.html]{2006astro.ph..2514Olsen},
and the mid-infrared data with Spitzer/IRAC at 3.6, 4.5, 5.6 and
8.0$\mu$m (Dickinson et al., in prep. and
http://data.spitzer.caltech.edu/popular/goods/, see also brief
descriptions in \citealp{2004ApJ...616...63Yan} and
\citealp{2006astro.ph..4250Stark}).

While we used the publicly-available reduced images for the ACS and
ISAAC data, we performed post-pipeline processing on the
GOODS-S/Spitzer data (from epochs 1 and 2) creating deep
1\arcmin\,$\times$1\arcmin\ mosaics centered on each high redshift
candidate using the Spitzer/MOPEX package. The resultant drizzled
mosaics have had instrumental effects and cosmic ray events
removed. The mosaics were registered to the ACS astrometric reference
frame using common detections to an accuracy of $\sim$0.1''. We
verified our IRAC mosaic generation and photometric calibration by
performing the same procedure to extract photometry for A and F stars
identified in the field \citep{2002A&A...392..741Groenewegen}. Such
stars rarely show an infrared excess or have strong absorption
features in the mid-infrared, thereby enabling accurate predictions of
their mid-infrared emission. The fluxes predicted by black-body fits
to the templates of their spectral types were reproduced to within
0.1, 0.1, 0.2 and 0.3\,mag of our aperture photometry at 3.6, 4.5, 5.8
and 8$\mu$m, respectively, consistent with our adopted accuracies (see
later for a description of our adopted photometric uncertainties).
Additionally, we confirmed that our measured fluxes were consistent
with those measured for our LBG candidates and the stars from the
epoch 1 and 2 GOODS Enhanced Legacy products (Dickinson et al., in
prep. and http://data.spitzer.caltech.edu/popular/goods/).

The deep GOODS/IRAC data have the largest spatial pixels and PSFs
amongst our datasets and, as a result, in crowded regions the emission
profile from a given source can overlap with those of adjacent
sources.  For those sources selected according to the above criteria,
but blended with neighbours in the Spitzer data, we treated the
mid-infrared IRAC fluxes as upper limits in the subsequent SED
analysis.

For each source that satisfied our Lyman break criteria, we built a
10-band SED with fixed diameter aperture photometry performed on each
image (see Figure \ref{fig:seds}). Apertures, appropriate for the
characteristics of each image, were determined based on enclosing as
large a fraction of the source flux as possible while minimising the
impact of the sky background noise and flux contributions from
neighbouring sources.  We used apertures with diameters of
$1^{\prime\prime}$, $2^{\prime\prime}$, and $4.5^{\prime\prime}$ for
the measurements on the ACS, ISAAC, and IRAC images, respectively.
Small aperture correction factors were applied to account for the
fraction of the total flux not enclosed by our fixed-diameter
apertures: 1.07 for the four ACS bands, 1.10 for the two ISAAC bands,
1.18 for the IRAC 3.6 and $\rm 4.5~\mu m$ bands, and 1.35 and 1.47 for
the 5.8 and $\rm 8~\mu m$ bands.  We determined the aperture
corrections based on the point spread function (PSF) of each mosaic,
constructed from isolated, bright but unsaturated stars throughout the
CDF-S and taking reference total apertures of diameters
$6^{\prime\prime}$ for the optical and near-infrared data, and
$12.2^{\prime\prime}$ for the mid-infrared data.

The formal uncertainty assigned to each photometric data point
was derived from measuring the background fluctuations in the parent
image using $\sim$10$^{3}$-10$^{4}$ synthetic apertures with
  fixed diameters laid down at random on regions that are empty of
flux from discrete sources \citep[\rm{e.g.}][]{2006AJ....131.1891Foerster},
thereby accurately characterising the photometric uncertainty directly
from the data, rather than calculating it
from the pixel-to-pixel rms assuming uncorrelated Gaussian noise
statistics.  The simulations were performed for a range of aperture
diameters.  As a function of linear aperture size $N$, the empirically
derived relationship shows that the photometric uncertainties lie above
and grow faster with size than the $\sigma \propto N$ dependence for pure
Gaussian noise.  However, for a given aperture size, the histogram of
background fluxes is very well described by a Gaussian distribution
(of which the dispersion is taken as the $1\sigma$ uncertainty for
photometric measurements in the given aperture size).
A conservative absolute calibration uncertainty of ten per cent was
additionally included in the final adopted uncertainty to
comfortably account for the relative uncertainties in photometric
calibration across the ten bands.  
In our SED modelling, this additional
uncertainty prevents the fits being driven by a few photometric points
with very small errors.
We note that adopting a less conservative value for this uncertainty
would mainly act to reduce the range of acceptable older ages, which is
most influenced by the IRAC photometry, and therefore would not alter our
findings of young ages for the majority of our sources.

\subsection{Defining a robust sample of \zfv\ Lyman-break galaxies}

Our initial rest-frame UV selection yielded a sample of 109 \zfv\ LBG
candidates. This sample will include a fraction of spurious sources
that are not LBGs (low redshift galaxies, quasars and stars) and also
some sources with insufficient photometric constraints (regardless of
their nature) to reliably determine their properties. We have
therefore culled the inital sample as follows.

Assisted by SEDs extending to the observed-frame mid-infrared, we have
reliably excluded 22 candidates from our sample which are low redshift
galaxies (or interlopers). Synthesis modelling of the colour evolution
of stellar populations predicts that galaxies with z$\,<\,$4 can also
satisfy our UV selection criteria.
These galaxies have intrinsically redder near and
mid-infrared SEDs than true high redshift Lyman break galaxies.
The SED of an example interloper is shown in panel (vii) of
Fig. \ref{fig:seds} and is easily distinguished from the SEDs of high
redshift LBGs (panel ii-vi).  Furthermore, Figure \ref{fig:seds}
demonstrates how the addition of the IRAC data has greatly enhanced
our ability to screen for such low redshift galaxies which are
generally the brightest sources in the IRAC wavelengths that satisfy
our optical selection criteria.  These interlopers mostly satisfy the
criterion for being extremely red objects \citep[ERO,
  $i_{AB}-K_{s,AB}>$2.48,][]{2003MNRAS.346..803Roche} and
IRAC-selected EROs \citep[iERO, $m_{3.6\mu m,AB}-
  z_{AB}>$20,][]{2004ApJ...616...63Yan}, consistent with being
galaxies at z$\,\sim\,$1-2. The importance of this screening is clear
when one considers that most samples of systems satisfying the
high-redshift LBG selection criteria distinctly lack confirmatory
spectroscopy and several of the recently reported IRAC-detected
modelled z$\,>\,$5 LBGs do not have confirmed spectroscopic redshifts
\citep{2005ApJ...618L...5Egami,2005ApJ...635..832Mobasher,2005ApJ...634..109Yan}.
In the case of the z$\,\sim\,$6 LBG in
\citet{2005ApJ...635..832Mobasher}, an interloper solution is more
plausible \citet{2006astro.ph..6192Dunlop}.

Our initial optical selection potentially includes low mass stars
within the Galaxy and QSOs. These are spatially unresolved in the ACS
data are therefore straightforwardly excluded. We identified 19 such
objects.  These objects generally have the brightest visible emission
amongst the systems that satisfy our selection criteria ($z_{AB} \sim
$23). \cite{2006astro.ph..1367Vanzella} obtained spectra of four of
these objects with sufficient quality for them to be classified as
stars, supporting the exclusion of unresolved objects from our robust
sample. Had we included them and subjected them to synthesis
modelling, their best-fit models would imply implausibly high star
formation rates and young ages. Thus, without screening for these
objects, our derived ensemble properties would be heavily biased
towards young ages and high star formation rates.

Finally, as accurate estimates of the ages, masses and star-formation
histories of our galaxies are dependent upon well-sampled SEDs, it is
imperative to analyse the objects with the most robust photometry.
Therefore, we additionally excluded the results from 47 sources with
secure detections in only two or three bands (including $i$ and $z$),
with the remaining constraints being upper limits. There are
insufficient data to constrain the models to accurately constrain
their physical properties, and these systems are flagged as having
'insufficient photometry'. A significant fraction of these systems are
blended with a nearby source in the IRAC bands. Because blending is
purely a random alignment of foreground sources with our high redshift
candidates, there is no reason these should be different to the
isolated candidates. Accordingly, we do not find any evidence for a
systematic difference in the derived properties of the blended and
isolated sub-samples. We note that this step may preferentially
exclude LBGs with high star formation rates, low masses and young ages
as these are less likely to be detected in the IRAC bands. Massive, old
LBGs selected at the same $i-$band magnitude as younger, lower mass
systems are intrinsically brighter in the observed IR and so require
more of the photometric bands to be compromised in order to exclude
them in this way.

\begin{figure}
\includegraphics*[width=8.5cm]{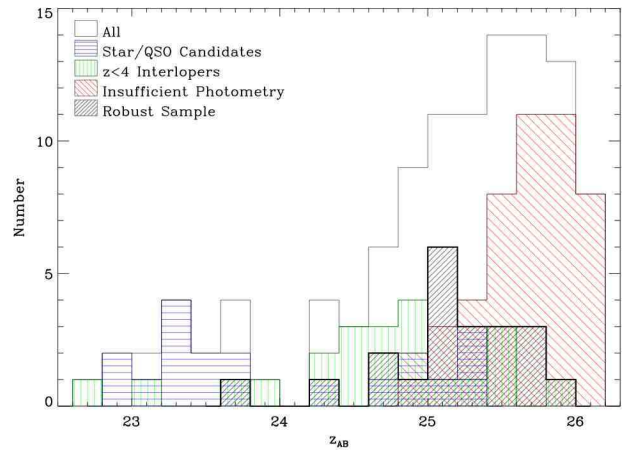}
\caption{Histogram showing the $z$-band magnitude distributions of the
  full set of 109 objects satisfying the initial selection criteria,
  with histograms of the comprising sub-samples overlaid. The
  black/shaded histogram represents the final 'robust' sample of
  reliable LBG candidates.}
\label{fig:histo}
\end{figure}

This culling process renders a final robust sample of 21 galaxies
which is the focus of our study.  Since each step in the culling
process (apart from the last) is unbiased with respect to the physical
properties of our high redshift galaxies, the final robust sample is
expected to be representative of the bright, \zfv\ galaxy population
selected by the Lyman-break technique, with the possibility that the
lower mass and younger systems may be underrepresented.  The use of
the longer wavelength near- and mid-infrared data have enabled us to
construct a uniformly selected rest-frame UV sample that is
reliable. Figure \ref{fig:histo} shows the magnitude distribution of
the robust sample in comparison to all objects satisfying our
selection criteria and the culled members. As expected, we find that
the Star/QSO candidates and low-redshift galaxies are in the most part
the brightest sources that satisfy our initial selection criteria. The
histograms of the robust sample and the LBG candidates flagged as
having insufficient-photometry suggest that the former comprises the
brighter members of the \zfv\ population and the latter the fainter
members.  Indeed 6 LBGs from the robust sample, and 9 candidates with
insufficient photometry, have been spectroscopically confirmed to be
\zfv\ LBGs \citep[see Section
  \ref{spectroscopy},][]{2005A&A...434...53Vanzella}.

\section{Evolutionary Stellar Synthesis Modelling}
\label{model}

Using a library of synthetic spectra
\citep{2003MNRAS.344.1000Bruzual}, we modelled the SED of each LBG
candidate simultaneously deriving photometric redshifts and the key
properties of age, extinction, star formation rate and stellar mass.
We explored a range of star formation histories (constant star
formation rate, instantaneous burst, and exponentially decaying star
formation rates with $e$-folding timescales ranging from 10\myr\ to
1\gyr).  The UV emission from galaxies is dominated by the population
of short-lived massive OB and A stars.  Models of continuous star
formation enable the production of these massive stars even at old
ages.  In contrast, massive stars die away rapidly in an instantaneous
burst model and conspicuous UV emission is only produced at very young
ages.  For a given initial mass function (IMF), model fits with these
two star formation histories will bracket the possible age range of
the galaxies.  Our emphasis on the constant star formation model
therefore provides upper limits to the best-fit stellar ages.  For
this reason we concentrate our analysis on the results obtained for a
constant star formation rate. The effects of adopting alternative star
formation histories on our results are discussed in Appendix
\ref{alt}.

For all of our models, we used a Salpeter stellar IMF between 0.1 and
100\msun.  The actual IMF is unconstrained for our objects.  A steep
rise down to the lower mass cutoff likely is unrealistic in view of
the turnover below 1\msun\ inferred for the local IMF
\citep[\rm{e.g.}][]{2001MNRAS.322..231Kroupa,2003PASP..115..763Chabrier},
and recent comparisons of dynamical masses and photometric stellar
masses suggest that Kroupa/Chabrier-type IMFs would be more
appropriate at higher redshift as well
\citep[\rm{e.g.}][]{2006ApJ...645.1062Foerster}.  The main impact of
changing the IMF on our SED modelling is on the derived masses and
star formation rates which, for a given rest-frame V-band luminosity,
would be about a factor of 1.4-2 lower with the
\citet{2001MNRAS.322..231Kroupa} or \citet{2003PASP..115..763Chabrier}
IMFs (\citealp[see][]{2003MNRAS.344.1000Bruzual}; \citealp[see also
  the discussion by][in the context of SED modelling of
  \zth\ LBGs]{2001ApJ...559..620Papovich}).  At z\,$\ga$\,2, and more
so at \zfv, top-heavy IMFs with a larger proportion of massive ($\rm >
10$\msun) stars, or even extremely massive ($\rm > 100$\msun)
metal-free stars formed from primordial gas (``population III'' stars)
may be an issue
\citep[\rm{e.g.}][]{2006MNRAS.369..825Schneider,2004ApJ...605..579Yoshida,2005MNRAS.356.1191Baugh}.
If so, using such an IMF in the modelling would lead to a reduction of
the derived stellar masses and SFRs.

We adopted \citet{2003MNRAS.344.1000Bruzual} models with a metallicity
of one-fifth solar and, for consistency, a Small Magellanic Cloud-type
(SMC-type) dust extinction law
\citep{1984A&A...132..389Prevot,1985A&A...149..330Bouchet}.
Observational constraints on the metallicity of \zfv\ galaxies are
scarce; however, the metallicities derived from absorption lines in
the spectra of eigth \zfv\ LBGs in the Subaru Deep Field are $\rm
Z~\sim~0.2~Z_{\odot}$ \citep{2004ApJ...610..635Ando}, suggesting that
our choice of $\rm 0.2~Z_{\odot}$ is reasonably representative.  For
each individual galaxy, best-fit parameters were derived from
minimisation of the reduced $\chi^{2}$ statistic.  
All model fitting followed standard procedures applied in similar studies  
of high redshift galaxies \citep[\rm{e.g.},][]
{2001ApJ...559..620Papovich,2006ApJ...640...92Papovich,2001ApJ...562...95Shapley,2005ApJ...626..698Shapley,2004ApJ...616...40Foerster,2006astro.ph..4554Yan}. 
We assumed a constant marginalisation for each of the
models.
While our results are marginalised over the extinction law, metallicity,
star formation history and IMF, we have actually explored a range of all
of these parameters (see Appendix \ref{alt}) but the results presented   
correspond to our preferred single values of these properties as described
above. Detailed aspects of the modelling procedure 
will be described in a forthcoming paper (N.M. F\"orster Schreiber et al.,
in preparation, Paper II).

\begin{figure}
\begin{centering}
\includegraphics*[width=7cm]{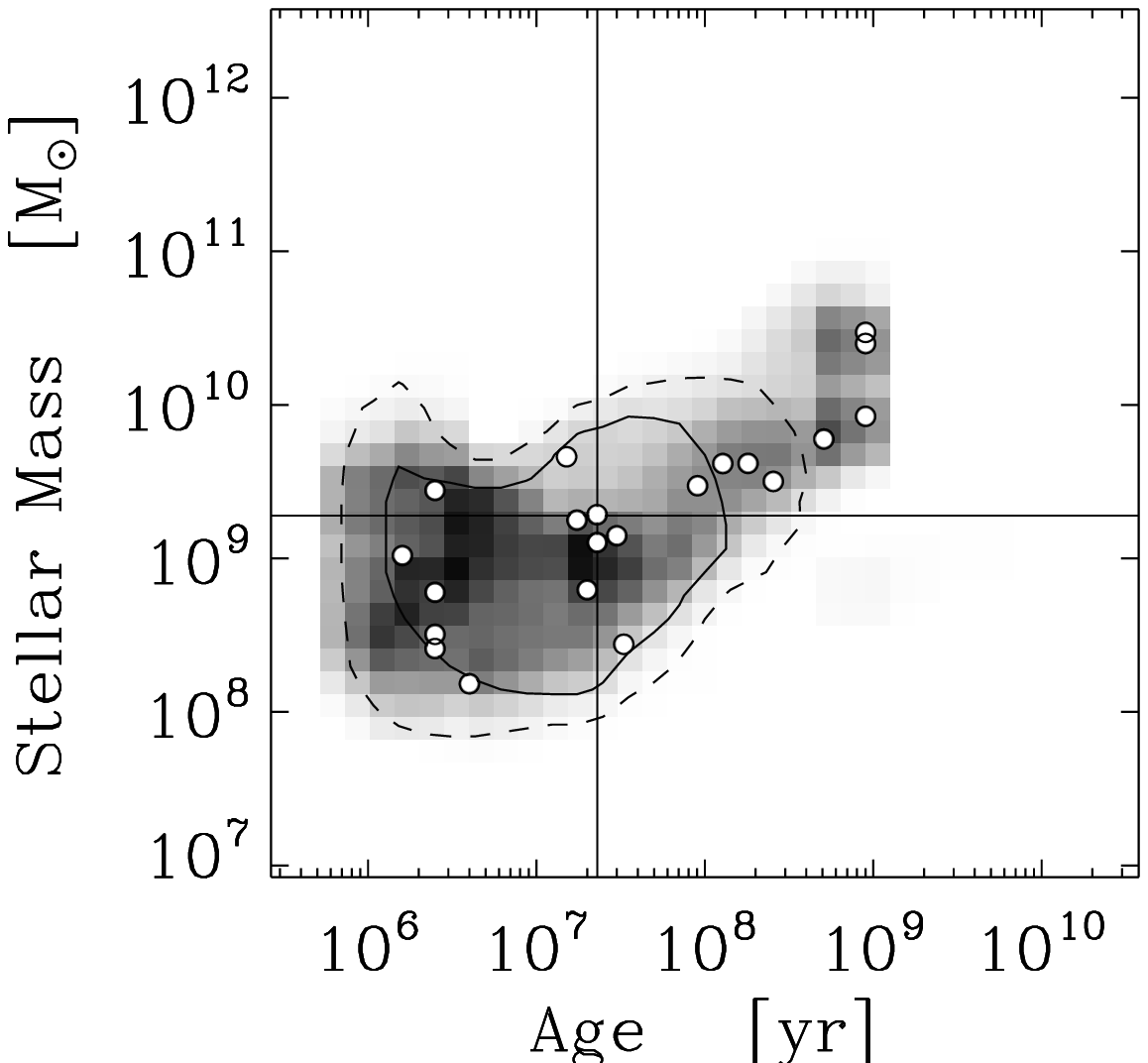}
\includegraphics*[width=7cm]{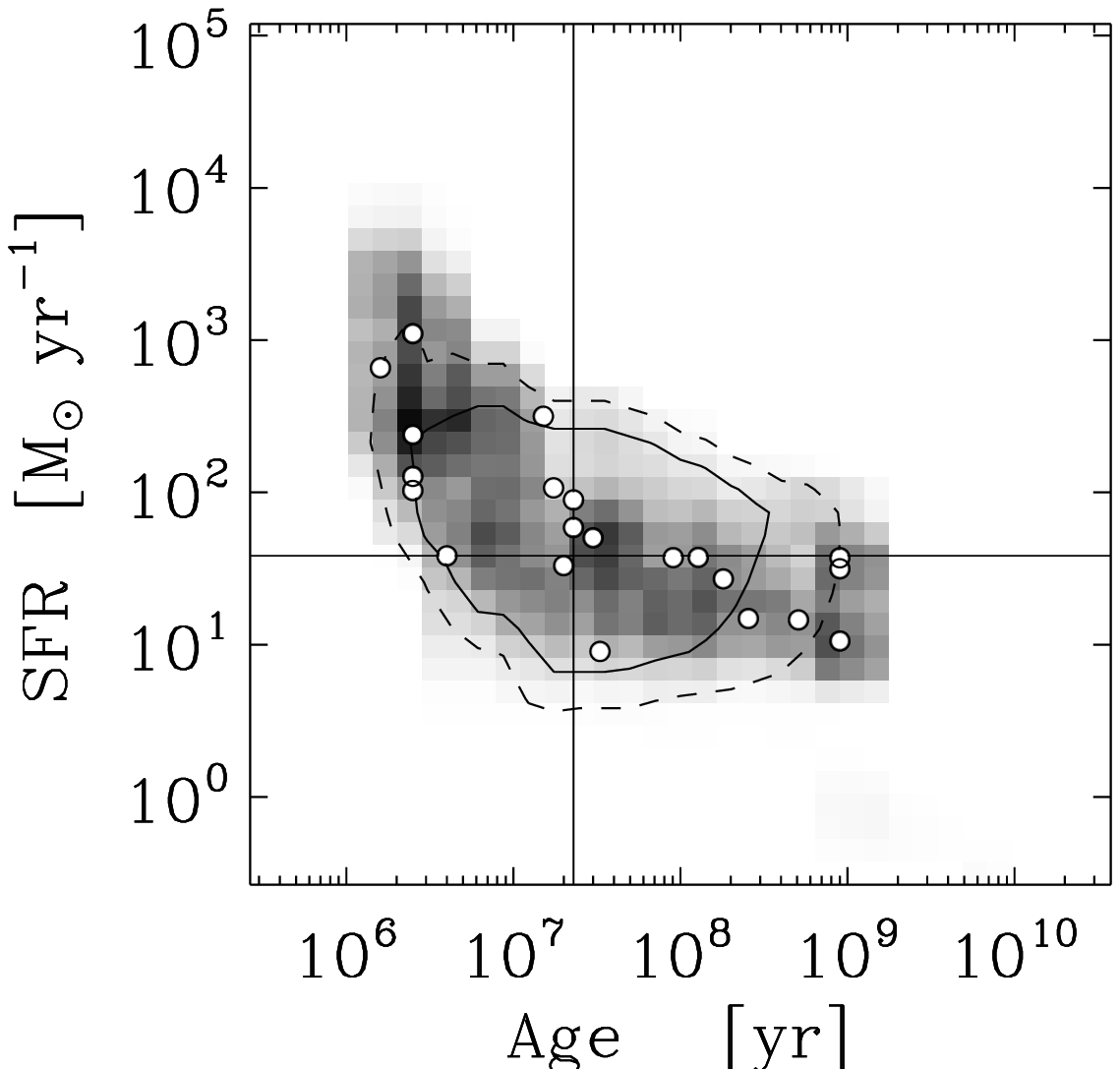}
\includegraphics*[width=7cm]{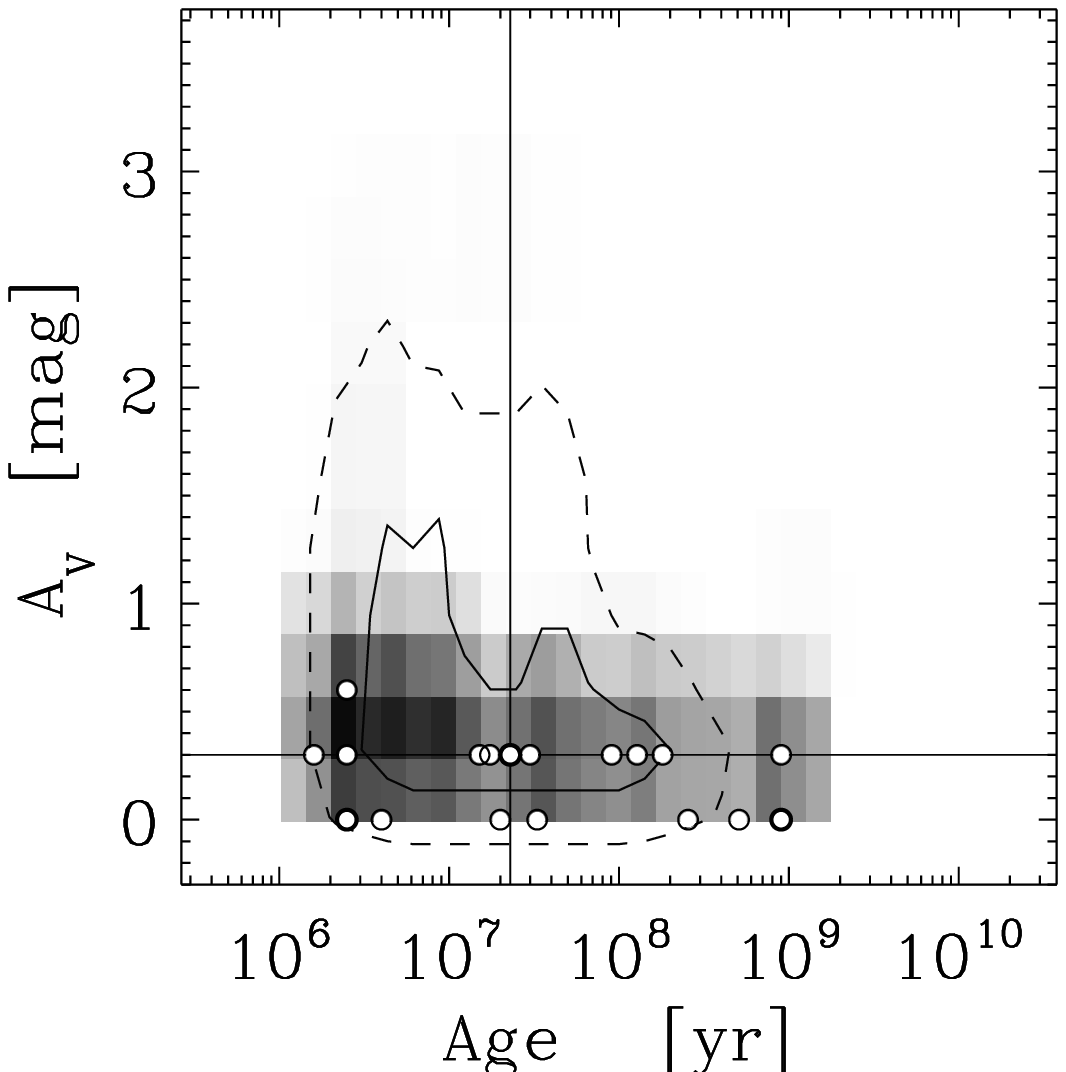}
\end{centering}
\caption{Composite probability distributions of age and (a) stellar
  mass, (b) star formation rate and (c) extinction for the robust
  sample of \zfv\ LBGs as determined for each galaxy from our 500
  Monte Carlo simulations. The points overlaid show the best-fit
  properties for each object in the sample and the large cross
  indicates the formal best-fit median properties of the entire
  ensemble. The overlaid contours indicate the 68\% (solid line) and
  95\% (dashed line) integrated probabilities on the ensemble
  properties measured from the centroid of the distribution. In other
  words, 68\% (and 95\%) of all sources should have ages and stellar
  masses that fall within these regions. The bulk of the composite
  probability distribution lies at ages less than 100\myr, as do the
  best-fit ages for the majority (two-thirds) of the sources.}
\label{fig:pd}
\end{figure}

We note that it is difficult to differentiate between the exact
metallicity, extinction by dust and star formation histories for each
individual galaxy based solely upon the goodness-of-fit because of
degeneracies among these model parameters.  Therefore, to determine
the confidence intervals for the modelled properties, we ran 500 Monte
Carlo simulations for each object.  We applied the same best-fitting
procedure after perturbing the input broadband photometry assuming the
photometric uncertainties are Gaussian as indicated by our background
noise fluctuations analysis (see Section \ref{sample}).  The results
of these simulations provide the probability distribution in parameter
space for each source.  By combining these individual probability
distributions, we derived those for the properties of the ensemble of
sources \citep[see, \rm{e.g.},][for an analogous
  approach]{2001ApJ...559..620Papovich}.  Example combined probability
distributions for the age, stellar mass, star formation rate and
extinction for all of the galaxies in the robust sample are shown in
Figure \ref{fig:pd}.  These diagrams show the probability that any
galaxy satisfying the Lyman-break criteria has of having those
properties. From these we can therefore determine the characteristic
properties of the ensemble of galaxies without relying solely upon the
mean of the (degenerate) best-fit solutions.

\section{Ensemble properties of the robust sample of \zfv\ LBGs}
\label{results}

\subsection{Redshifts: Photometric and spectroscopic}
\label{spectroscopy}

\begin{figure*}
  \includegraphics*[width=7.5cm]{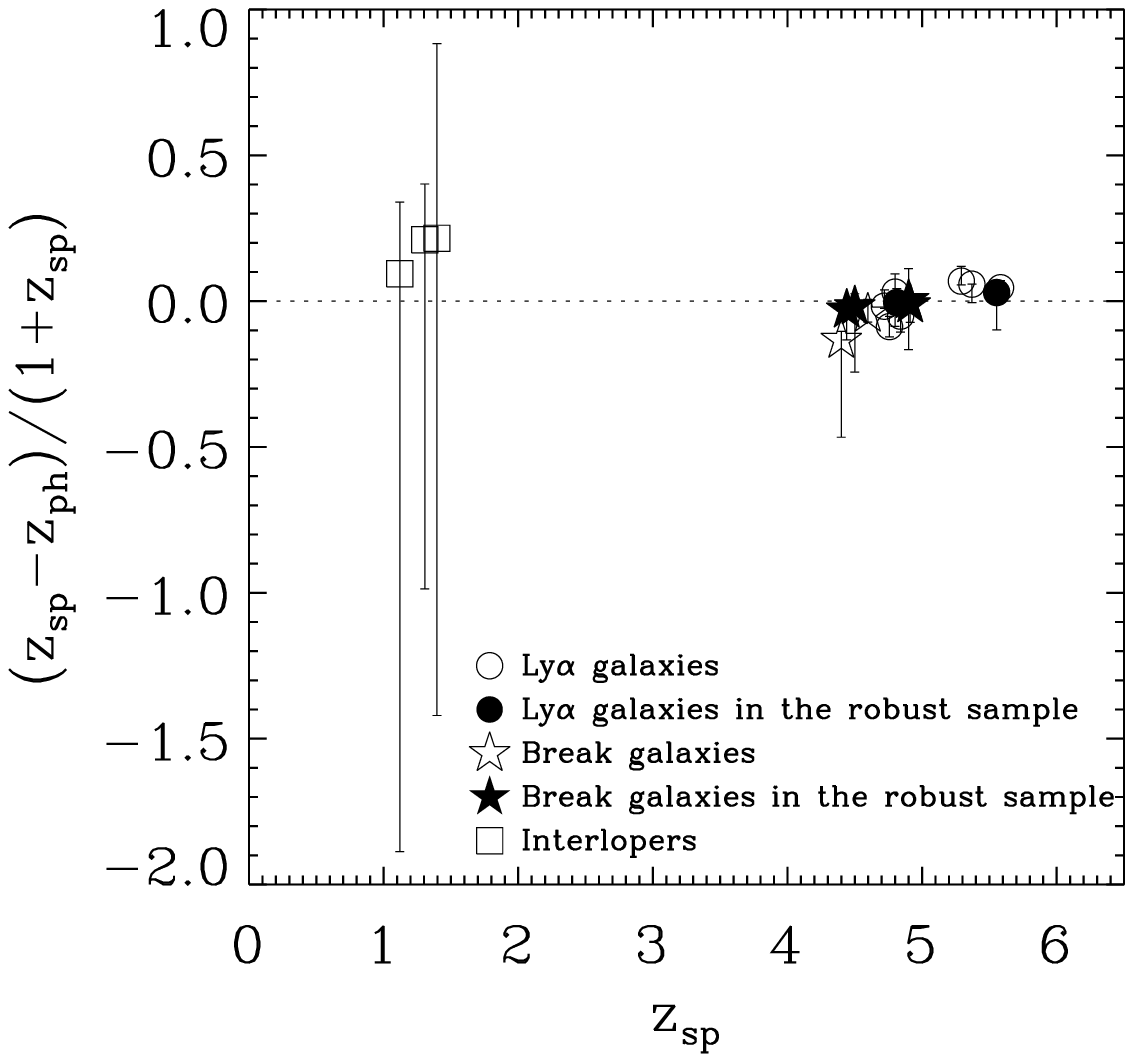} 
  \includegraphics*[width=8.5cm]{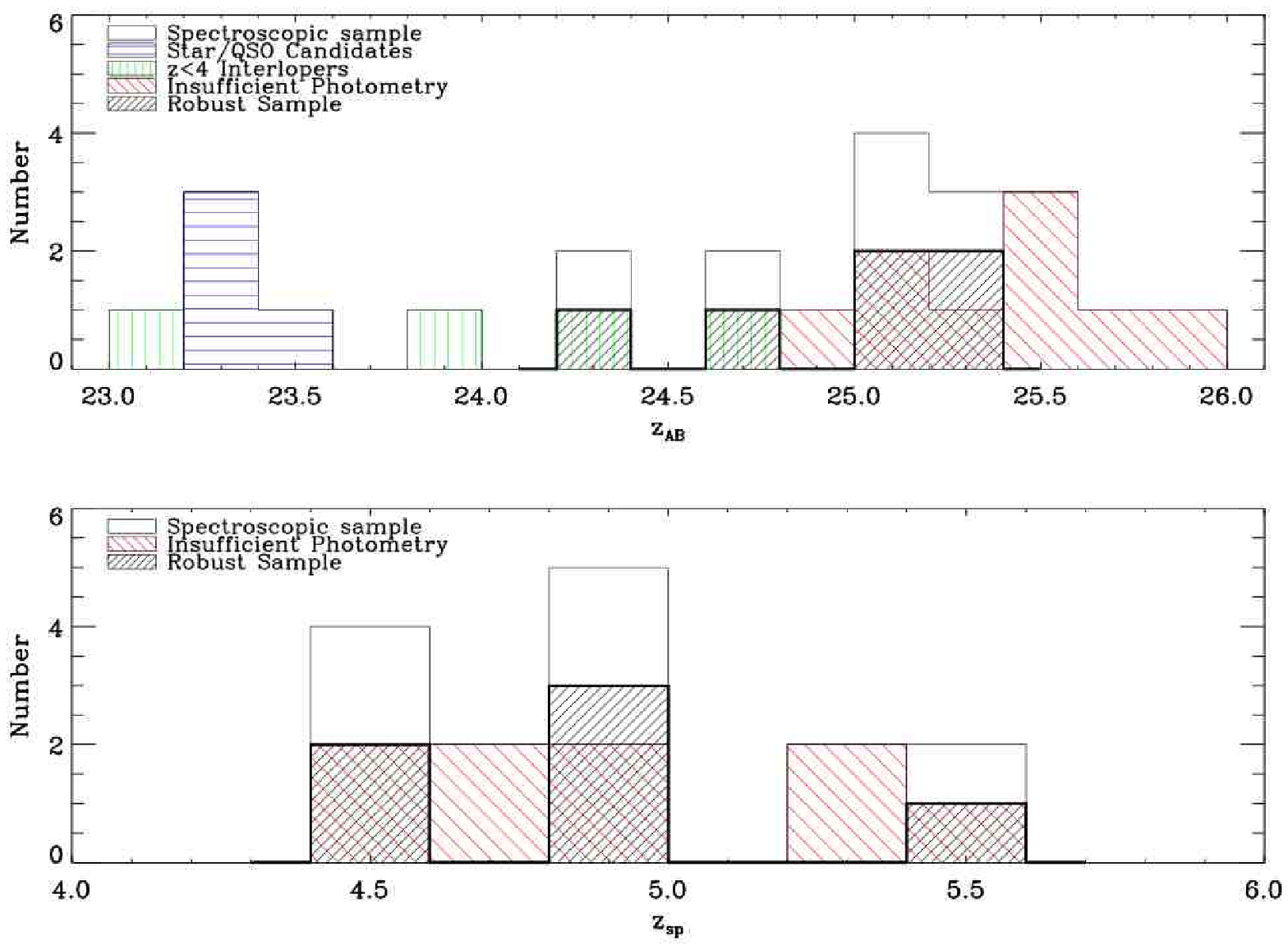} 
  \caption{Left: This figure shows the agreement between
    spectroscopic and photometric redshifts of all objects which
    satisfy our selection criteria which are confirmed to be
    galaxies. All filled symbols are for \zfv\ LBGs from our robust
    sample. Two $\sigma$ error bars on the photometric redshift are
    overplotted.  The figure excludes a low redshift interloper which
    has an x-ray counterpart (XCDFS265) and is likely to be
    AGN-hosting, therefore the results of the stellar evolutionary
    synthesis modelling does not make sense in this case. As a guide,
    the line z$_{sp}$=z$_{ph}$ is overplotted. Right: $z$-band
    magnitude distribution of the spectroscopically confirmed LBGs. As
    in Figure \ref{fig:histo} the black/shaded histogram corresponds
    to galaxies in the 'robust' sample.}
\label{fig:zspzph}
\end{figure*}

The LBGs in the robust sample have best-fitting photometric redshifts
in the range 4.60 to 5.54 ($<$z$>$$_{median}$ = 4.8) as expected given
our initial photometric selection\footnote{Using standard procedures
  \citep{1996MNRAS.283.1388Madau} we estimate the redshift range our
  selection criteria are sensitive to. This range is determined
  following the implementation in
  \citet{2003ApJ...593..630Lehnert}. Standard local galaxy templates
  were modified to account for IGM opacity along the line of sight
  short-ward of Ly$\alpha$ \citep{1996MNRAS.283.1388Madau}. They were
  then further modified using the redshift dependent Gunn-Peterson
  absorption of Ly$\alpha$ as described in
  \citet{2005astro.ph.12082Fan}. This results in star-forming galaxies
  lying between z$\sim$4.6 and z$\sim$6 being able to satisfy our
  selection criteria.}. Our photometric redshifts agree exceptionally
well (within 2$\sigma$, or better), for both the high redshift LBGs
and low-redshift interlopers, with spectroscopic redshifts where
available (see Figure \ref{fig:zspzph} and Table \ref{tab:zspzph}).

As part of the public spectroscopic surveys of the CDFS conducted by
ESO \citep[but also see v1.2 release
  http://www.eso.org/science/goods/spectroscopy/CDFS\_Mastercat/]
{2005A&A...434...53Vanzella,2006astro.ph..1367Vanzella}, spectroscopy
has been performed for 7 LBGs from our robust sample, 6 of which are
confirmed to be at \zfv. The low signal-to-noise ratio of the spectrum
of the remaining robust LBG prevented assignment of a redshift. This
high confirmation rate reinforces that our selection of \zfv\ LBGs is
robust. Spectroscopic redshifts were also secured for 17 additional
sources: 4 are stars, 4 are low-redshift interlopers (all agreeing
with our photometric/morphological classification) and 9 are confirmed
to lie at \zfv\ and are part of our 'insufficient photometry'
sub-sample which we expect includes genuine \zfv\ LBGs. The right
panel of Figure \ref{fig:zspzph} shows the $z$-band magnitude and
redshift distributions of the spectroscopically confirmed redshift 5
LBGs from the robust sample and those flagged as having insufficient
photometry.  The latter are slightly fainter than the robust sample
while the redshift distribution is similar indicating \textit{bona
  fide} fainter LBGs are among the sources flagged as having
insufficient photometry.

\begin{table*}
\caption{Breakdown of photometric and spectroscopic redshifts for all of the LBG candidates. The z$_{phot}$ column refers to the nominal best-fit redshift determined from our modelling and the z$_{spec}$ to spectroscopic redshifts from the literature.}
\label{tab:zspzph}
\begin{centering}
\begin{tabular}{@{}lccccccccc@{}}
  \hline 
& & \multicolumn{3}{c}{z$_{phot}$} & & \multicolumn{4}{c}{z$_{spec}$} \\ 
 & & Total & z$<$4 & z$>$4 & & Attempted & z$<$4 & z$>$4 & Unassigned\\ 
 & & (1a)  & (1b)  & (1c)  & & (2a) & (2b) & (2c) & (2d) \\
\hline 
UV Selection & & 109 & 23 & 86 & & 34 & 8 & 15 & 11 \\ 
\hline
 Robust sample & & 21 & 0 & 21 & & 7 & 0 & 6 & 1 \\ 
Insufficient Photometry & & 47 & 1 & 46 & & 14 & 0 & 9 & 5\\ 
Star/QSO & & 19 & 0 & 19$^\dagger$ & & 5 & 4 & 0 & 1\\ 
Interlopers & & 22 & 22 & 0 & & 8 & 4 & 0 & 4\\ 
\hline
\end{tabular}
\end{centering}

\medskip
\raggedright Columns 1(a-c) refer to the results of the SED modelling for all candidates (109) that satisfy our original UV-selection, and columns 2(a-d) for those of the 109 candidates which have been spectroscopically follow-up.\\
(1a) Total number of sources satisfying our initial selection criterion.\\
(1b) \& (1c) The numbers of LBG candidates with photometric redshifts above and below z=4.\\
(2a) Total number of LBG candidates from our initial sample that have had spectroscopic measurements.\\
(2b) Number of candidates for which the spectra were
of insufficient quality to definitively assign a redshift.\\
(2c) \& (2d)  The numbers of sources with confirmed spectroscopic redshifts above and below z=4.\\
$^\dagger$ Best-fit redshifts of z$_{phot}$$>$4 are produced for all
objects classed as stars/QSOs due to the the strong break in their
SEDs being identified with the Lyman-break and because their SEDs
are fit with inappropriate star-forming galaxy templates.\\
\end{table*}

\subsection{Stellar Masses, ages, star formation rates and extinction}

The rest-frame SEDs of the majority of our prime candidates are very
blue from the far-UV to the visible, as is expected for young stellar
populations with no more than moderate dust obscuration. The results
of our stellar evolutionary synthesis modelling with the parameters
specified in Section \ref{model} yields the following typical
properties for the galaxies. These are typically strongly star forming
galaxies with a median star formation rate of 40\sfr. The median
photometric stellar mass is 2$\times$10$^{9}$\msun, about a factor ten
lower than Lyman-break galaxies at
\zth\ \citep{2001ApJ...562...95Shapley,2001ApJ...559..620Papovich}.
The stellar mass estimates are the most robust of all derived
properties, varying by a factor of 2-5 depending on the model
assumptions (see Appendix \ref{alt}).  While a large spread in the
best-fit age is seen amongst individual sources in the sample the
median age is relatively young, $\sim$25\myr. The best-fit models
imply only moderate extinction with a median value in the V-band of
A$_{V}$ = 0.3 mag. Models including extinction at this level fit the
SEDs of the \zfv\ LBGs better than models without extinction.

As discussed in Section \ref{model}, we use the results of our
Monte-Carlo simulations to characterise the properties of the ensemble
of \zfv\ LBGs. Figure \ref{fig:pd} shows the combined probability
distributions for the parameters of mass, star formation rate and
extinction as a function of age of the stellar population.  While the
ensemble probability distribution in age extends over 1\myr\ to 1\gyr,
there is a clear concentration at young ages.  As indicated by the
contours shown in Figure \ref{fig:pd}, more than 68 per cent of the age
distribution lies at ages less than 100\myr, consistent with the
majority (two-thirds) of the galaxies having formal best-fit ages
younger than this. In addition, Figure \ref{fig:pd}(a) shows that
these young sources have stellar masses $<$10$^{10}$\msun\ and
typically $\sim$10$^9$\msun, and Figure \ref{fig:pd}(b) suggests they
are the most strongly star-forming systems within our sample. This is
unsurprising as they must have the lowest UV mass-to-light ratios in
our sample given their low mass.

The remaining third of our sample have best-fit masses of order
10$^{10}$M$_{\odot}$ and ages older than a few hundred million years.
These older galaxies are analogous to several galaxies recently
reported
\citep{2005ApJ...618L...5Egami,2005MNRAS.362.1054Schaerer,2005ApJ...634..109Yan,2005ApJ...635..832Mobasher,2005ApJ...635L...5Chary,2005MNRAS.364..443Eyles,2005ApJ...630L.137Dow,2006astro.ph..4554Yan}
that lie at similar or slightly higher redshifts (z$\,\sim\,$6-7) than
ours, where the presence of a discontinuity between the near and
mid-infrared bands in the SED is identified with the rest-frame
Balmer/4000\AA\ break (Figure \ref{fig:seds}).  This break is
indicative of more evolved stellar populations, where the emission
from stars with ages of several hundred million years dominates over
that from the short-lived massive stars that produce the UV emission.
The absence or weakness of a discontinuity between the $K_{s}$-band
and 3.6$\mu$m photometry points in the SEDs of the majority of our
sample of galaxies with younger best-fit ages strongly limits the
possible contribution to the integrated stellar light by such
similarly evolved stellar populations. Indeed, the lack of this
discontinuity constrains their best-fit ages to less than $\sim$100
million years. While old and massive systems are present in the
sample, our results clearly indicate that a substantial fraction
($>$two-thirds) of galaxies at \zfv\ satisfying the Lyman break
selection criteria are dominated by a young, intensely star-forming
component.  Similarly young ages and moderate masses (ages
$\la$45\myr\ and stellar masses $\sim$10$^9$M$_{\odot}$) are found for
IRAC undetected z$\,\sim\,$6 $i$-band-dropouts (z$\,\sim\,$6) LBG
candidates recently reported by \citet{2006astro.ph..4554Yan}.

In this section we have reported on the typical physical properties
for the sample of luminous \zfv\ LBGs for our adopted modelling
assumptions. We have extensively investigated the effects of varying
the input model parameters on the derived properties and find the key
properties of young ages and moderate masses are robust under a wide
range of model assumptions. We discuss the effects of varying the
input parameters in the Appendix (see also N.M. F\"orster Schreiber et
al. in preparation).

\section{Comparison to LBGs at redshift 3}
\label{comp}
\subsection{Stellar Masses, Ages, Star-formation Rates and Extinction}

\begin{figure*}
  \includegraphics*[angle=90.,width=17cm]{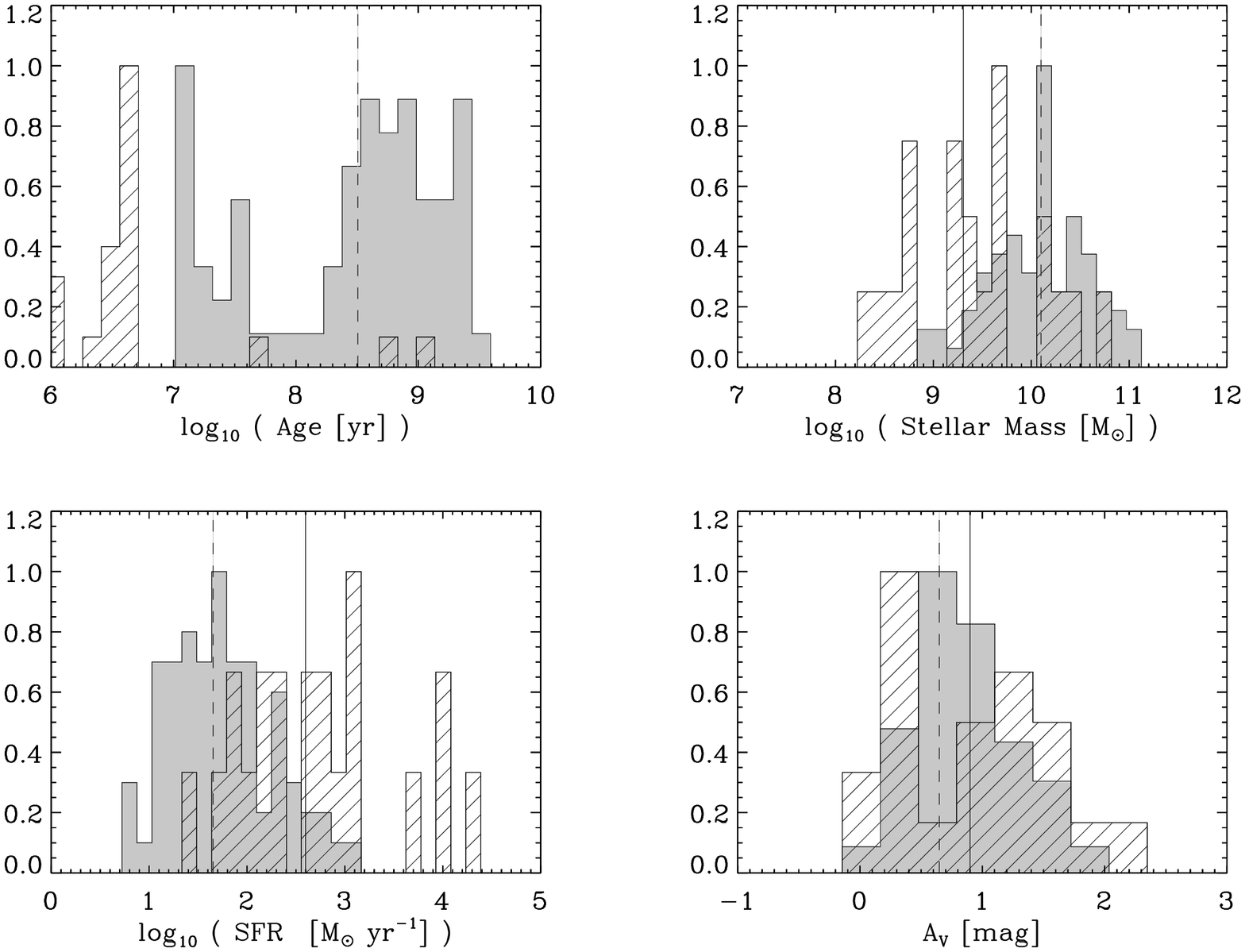}
\caption{Histograms (peak normalised) showing the distribution of
  best-fit properties of 74 \cal{R}$<$25.5 \zth\ LBGs from
  \citet{2001ApJ...562...95Shapley} (shaded grey) together with the
  best-fit properties of our robust sample of \zfv\ LBGs (line filled
  histograms). These properties correspond to identical model
  assumptions: constant star formation, 0.1-100\msun\ Salpeter IMF,
  Calzetti extinction law and solar metallicity templates of
  \citet{2003MNRAS.344.1000Bruzual}.  The vertical lines indicate the
  median values for the \zth\ sample (dashed line) and the
  \zfv\ sample (solid line). LBGs at \zfv\ are typically younger
  ($<$100\myr) and are less massive (few$\times$10$^9$M$_{\odot}$)
  than \zth\ counterparts of similar luminosity (see text for a
  detailed discussion).}
\label{fig:zsolshaphist}
\end{figure*}

\begin{figure*}
  \includegraphics*[angle=90.,width=17cm]{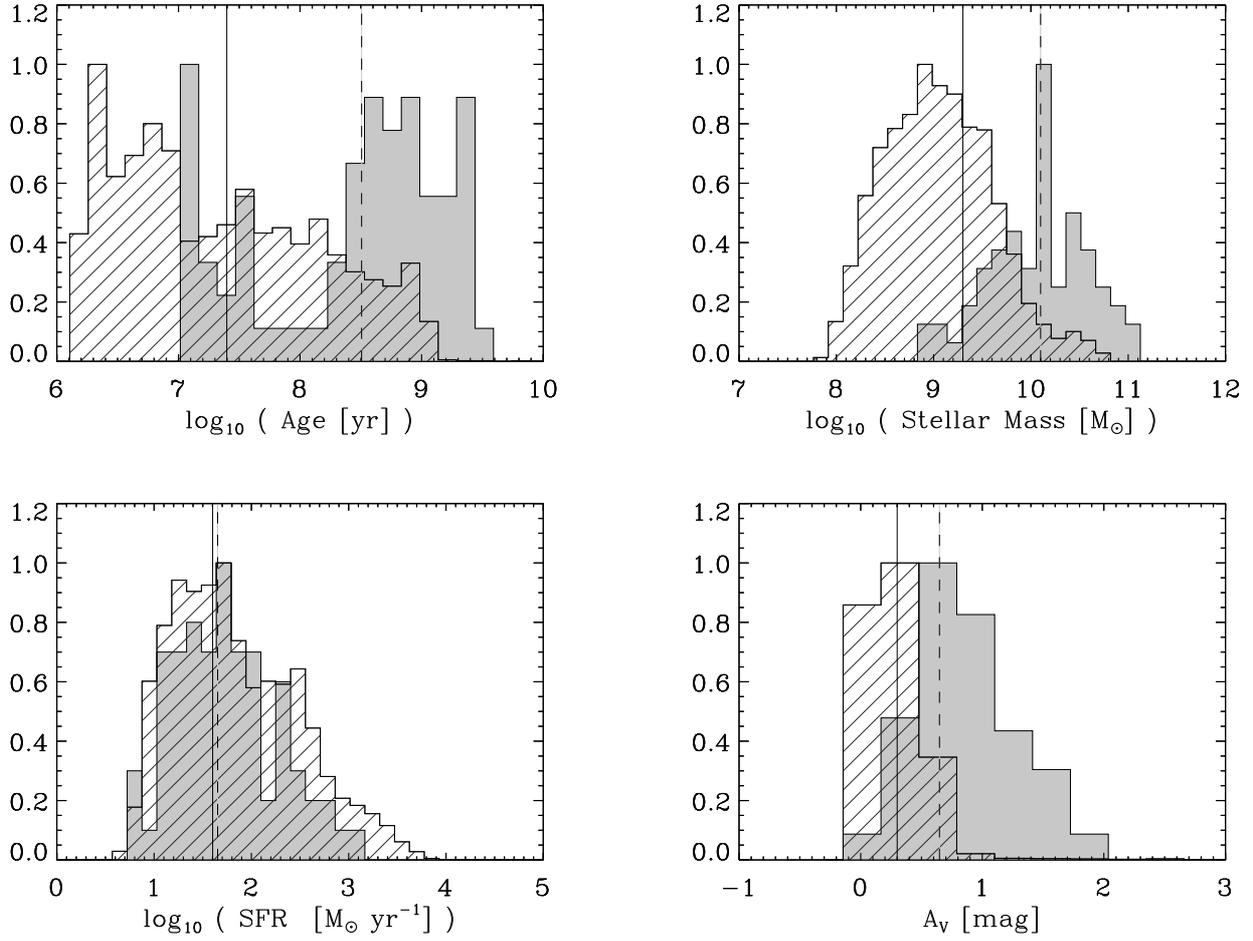}
\caption{The shading and lines are as in Figure \ref{fig:zsolshaphist} but now
  showing the distributions of properties derived from our Monte-Carlo simulations
  of the properties of \zfv\ LBGs from our robust sample (line filled histograms) obtained
  with 0.2\zsun\ templates and a SMC extinction law
  (see text for details).}
\label{fig:shaphist}
\end{figure*}

\begin{figure*}
  \includegraphics*[angle=90.,width=17cm]{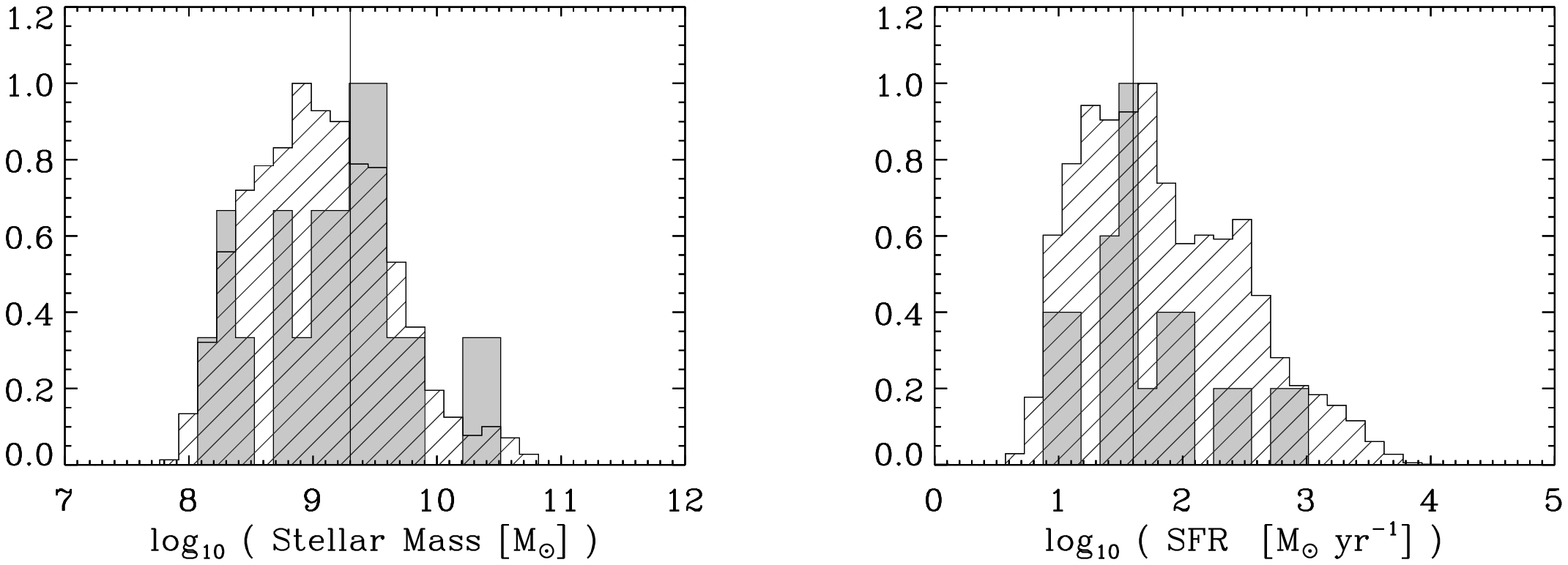}
\caption{This figure shows that the distribution of properties derived
  from our Monte Carlo simulations (line filled histograms) trace the
  distributions of nominal best-fit (shaded histogram) properties
  well. Stellar masses and star formation rates are shown, ages and
  extinctions show similar agreement. The solid vertical line
  indicates the median value.}
\label{fig:bfitshap}
\end{figure*}

In this section we compare the properties of our \zfv\ sample with
those published for the \zth\ LBG sample from
\citet{2001ApJ...562...95Shapley} which is matched in rest-frame UV
luminosity. The SEDs for both samples span the same rest-frame
wavelength range and modelling shows significant differences in the
stellar masses and ages derived even for identical modelling
assumptions. Because of this and the matched selection, we can only
ascribe these differences to intrinsic differences in the properties
of LBGs at these two epochs.

\vspace{3mm}

The $i$-band selection limit of $i_{AB}<26.3$ chosen for our
\zfv\ sample is matched to the typical magnitude limit of
spectroscopically confirmed samples of LBGs at \zth.  This limit
($\cal{R}_{AB}\,<\,$25.5), is sensitive to \zth\ LBGs that are $\sim$1
mag fainter than $m*_{1700\AA, AB}^{\zth}$=24.48
\citep{1999ApJ...519....1Steidel}. The $\cal{R}-$band maps to
1700\AA\ in the rest-frame at \zth.  This enables us to compare the
ensemble properties of our robust sample of \zfv\ LBGs to those
obtained for 74 \zth\ LBGs with apparent magnitudes greater than
0.1\textit{m*} from the sample analysed by
\citet{2001ApJ...562...95Shapley}.  As well as matching our samples by
magnitude, in order to make a fair comparison we have modelled the
SEDs of our robust sample of \zfv\ SEDs using the same assumptions as
\citeauthor{2001ApJ...562...95Shapley}. Specifically, we used solar
metallicity model spectra generated with the
\citet{2003MNRAS.344.1000Bruzual} population synthesis code, a
constant star formation history, a Salpeter IMF \footnote{They assume
  a slightly higher upper mass cut-off of 125\msun\ than our assumed
  100\msun, however this slight increase has an insignificant effect
  on the generated models and resultant properties.} and a Calzetti
attenuation law \citep{2000ApJ...533..682Calzetti}.  Note that this is
different a extinction law and metallicity than those used in previous
sections, but allows a direct comparison between the two samples. Any
differences in the results indicate intrinsic differences in the
properties of the two samples, either in properties like mass and star
formation rate, or that different extinction laws or metallicities are
required at the two epochs.

Figure \ref{fig:zsolshaphist} shows the result of this comparison for
the best-fit stellar masses, ages, extinctions and star formation
rates obtained for a constant star formation history from the
\zth\ LBG sample of \citet{2001ApJ...562...95Shapley} (shaded
histograms) and the results for our robust \zfv\ sample (line filled
histograms). While the derived extinctions are well matched, this
comparison clearly shows significant differences in the ages, masses
and star formation rates between the samples at \zth\ and \zfv. The
\zfv\ sample shows young typical ages ($\la$10\myr), in contrast the
\zth\ LBGs are significantly older ($\sim$320\myr). The typical
stellar mass of \zfv\ LBGs is $\sim$5-10 times lower than the
\zth\ sample (few\,$\times 10^{9}$\msun). Extremely high star
formation rates are estimated ($\sim$500\sfr) for the \zfv\ LBGs, a
factor of ten higher than the \zth\ sample.  Since we have compared
the properties of similarly selected galaxies, based on the same
rest-frame UV measurements, to the same UV luminosity, which have been
modelled under the same assumptions, these differences can
unequivocally be ascribed to inherent differences in the properties of
the two samples of LBGs.

Moreover, this comparison demonstrates that adopting Calzetti
extinction law and solar metallicity may not be appropriate for this
sample of bright \zfv\ LBGs.  Implausibly young ages ($<$10\myr) and
excessively high star formation rates ($>$10$^{3}$\sfr) are seen for a
large fraction of the \zfv\ population. The fact that the ages are so
signficantly younger than the \zth\ LBG population (see top left panel
Figure \ref{fig:zsolshaphist}) is an indication that the SEDs of
\zfv\ LBGs are even 'bluer' than their \zth\ counterparts. The
combination of the adopted extinction law and the high metallicity
templates results in these extremely young ages and high star
formation rates \citep[see the discussion in Appendix \ref{alt} and
  also][]{2001ApJ...559..620Papovich}. 

As discussed in Section \ref{model}, evidence suggests our consistent
choice of 0.2\zsun\ and the SMC extinction law is suitable to model
the blue SEDs of our \zfv\ LBGs.  \citet{2001ApJ...562...95Shapley}
have discussed the suitablilty of the Calzetti plus solar metallicity
models for \zth\ LBGs. In Figure \ref{fig:shaphist} we show the
comparison between the properties of LBGs at \zth\ and \zfv\ under
these appropriate (but different) assumptions. We have elected to show
the probability distribution of properties obtained from our Monte
Carlo simulations of the \zfv\ LBGs to give less importance to the
formal best-fit solution which can be sensitive to the modelling
assumptions.  Since the probability histograms follow the same
distribution as the best-fit parameters themselves (see example for
stellar mass and SFRs in Figure \ref{fig:bfitshap}), our comparison
using the probability distributions is reasonable.

Figure \ref{fig:shaphist} shows that the distributions of star
formation rates are extremely well matched, however, there are clear
differences in the distributions of stellar mass, age and extinction.
The typical stellar mass of LBGs is a factor of ten lower at redshift
5 than at redshift 3. Similarly, the distribution of ages shows more
sources have ages $<$100\myr\ in the \zfv\ sample than the
\zth\ sample.  There is an apparent difference in the distribution of
extinctions but this can largely be attributed to differences in the
modelling assumptions (Appendix \ref{alt}). This is corroborated by
the fact that the extinction distributions overlap almost entirely
when the same modelling assumptions have been made as we have shown in
Figure \ref{fig:zsolshaphist}. Furthermore, a comparison to the
results for similarly luminous \zth\ LBGs studied in
\citet{2001ApJ...559..620Papovich}\footnote{\citeauthor{2001ApJ...559..620Papovich}
  model \zth\ LBGs with the same metallicity
  \citep{2003MNRAS.344.1000Bruzual} templates and IMF as we have
  assumed. However, they give best fit results for individual galaxies
  only with the Calzetti extinction law. Since
  \citeauthor{2001ApJ...559..620Papovich} give best fit results for a
  range of star formation histories we consider only those fit with
  constant star formation and those with best-fit ages less than the
  e-folding time-scales of their 'best-fit' exponentially decaying
  star formation history model. The latter condition for age$<$$\tau$
  gives similar results to constant star formation. In addition, we
  considered only those LBGs from this faint sample which are matched
  in luminosity to the \zfv\ samples, V$<$24.8. Ten LBGs from
  \citeauthor{2001ApJ...559..620Papovich} satisfied these criteria.},
assuming a Calzetti extinction law but with the 0.2\zsun\ templates,
shows similar extinction for the \zth\ LBGs as is seen in Figure
\ref{fig:shaphist}. Thus the difference in the extinction between
\zfv\ and \zth\ LBGs seen in Figure \ref{fig:shaphist} is mostly a
consequence of the adopted extinction law and not the metallicity of
the templates.

\citet{2001ApJ...559..620Papovich} have comprehensively discussed the
effects of different model assumptions on the derived properties of a
faint sample of \zth\ LBGs. Their combined probability distributions
highlight the problems associated with using the Calzetti extinction
law with the low metallicity \citet{2003MNRAS.344.1000Bruzual}
templates. In their Figure 9b, they show the probability distribution
resulting from the SED modelling assuming a Calzetti extinction law
with the bluer 0.2\zsun\ metallicity templates. A clear extension of
sources with extremely young ages ($\la$~10\myr) and high $A_V$
($\sim$2) is seen. This extension almost disappears under a Calzetti
law with \zsun, and with the SMC law for either the 0.2\zsun\ or
1\zsun\ models (compare their figure 9b with 9d, 12b and 12d for a
Salpeter IMF).

Thus, adopting the Calzetti law with the solar or low metallicity
templates would only act to reduce the derived ages of the \zfv\ LBGs
and results in stellar masses that are 5-10 times lower than
\zth\ LBGs, \rm{i.e.} leaving our primary conclusions of young ages
and moderate masses for \zfv\ LBGs unaffected. Moreover, under all
suitable assumptions for the extinction law and consistent
metallicities (\rm{i.e.} 1\zsun\ with the Calzetti law, 0.2 or
1\zsun\ with SMC), we find that galaxies with ages ($<$100\myr) and
masses ($\sim 10^9$\msun) dominate samples of bright LBGs at \zfv. Our
analysis of the probability histograms and distributions shown in
Figures \ref{fig:pd} and \ref{fig:shaphist} shows that this fraction
to be $\ga$70 per cent.  However, while \zth\ LBGs with such young ages and
similar stellar masses are not unknown
\citep[\rm{e.g.}][]{2001ApJ...559..620Papovich,2001ApJ...562...95Shapley},
they are less than half as prevalent in LBG samples at \zth\ than at
\zfv\ ($\la$30 per cent).  This unequivocal change in the age and mass of the
\textit{dominant} population in luminous LBG samples separated by
$\sim$1\gyr, implies that at \zfv\ is an epoch where the majority of
LBGs are recently formed and are in the process of accumulating their
first significant mass.

\subsection{Duty cycle and bias}
\label{dutycycle}

We have shown in the previous sections that the dominant population of
LBGs in our \zfv\ sample is younger and of lower mass than that in
\zth\ LBG samples with similar UV luminosities. Since the majority of
LBGs at \zfv\ are so young, it is inevitable their detectability is
highly stochastic and this stochasticity impacts upon the
interpretation of other statistical properties such as their
clustering strength, number density and bias.

\citet{2003ApJ...593..630Lehnert} showed that the co-moving number
density of \zfv\ LBGs selected in a similar manner to our robust
sample was a factor of three lower than that of similar luminosity
LBGs at \zth\ and \zopen4 \citep{1999ApJ...519....1Steidel}.  If the
LBGs at both epochs had identical properties then this can be directly
related to evolution in the comoving number densities of the halos that
they occupy. Clearly, this na\"ive picture is complicated by the clear
difference in properties of the two samples in the previous
section. Moreover, the younger ages of the higher redshift sample
imply a shorter duty cycle for the LBG, related to the length of time a
galaxy would be detected as a UV luminous LBG.

Given that the average lifetime of the \zfv\ LBGs of $\sim$25\myr\ and
that the lookback time spanned by our survey (325\myr\ over the
redshift range of the \zfv\ LBGs 4.4\,$<$\,z\,$<$\,5.6), suggests
that luminous members of the \zfv\ LBG populations have a duty cycle
of 0.08. A similar estimate for \zth\ LBGs by
\citet{2001ApJ...562...95Shapley}, renders a duty cycle of 0.5,
significantly longer than at \zfv.  So across the two sample volumes,
the higher redshift sample represents systems $\sim$thirteen times as
numerous as those detected, whereas the lower redshift sample misses
only half of the potential LBGs in the volume. Clearly, the young ages
of bulk of the high redshift sample have a significant effect on the
bias of these sources.

LBGs have been long known to be a significantly clustered population
\citep[\rm{e.g.}][]{1998ApJ...505...18Adelberger,1998ApJ...503..543Giavalisco,2004ApJ...611..685Ouchi,2006ApJ...642...63Lee}\footnote{Further
  analysis on the clustering strength of \zfv\ LBGs will be presented
  in a forthcoming paper A. Verma et al. in preparation, Paper
  III}. Clustering analysis of 23.5\,$<\cal{R}_{AB}<$\,25.5 LBGs at
\zth\ suggests that LBGs reside in dark matter halos of masses
$10^{11.2-11.8}$\msun\ \citep{2005ApJ...619..697Adelberger} consistent
with recent estimates of the dynamical virial mass of the DM halos of
LBGs determined rotation curves of \zth\ LBGs
\citep{2005MNRAS.363L...6Weatherley,2006astro.ph..6527Nesvadba}.  For
\zfv\ LBGs, \citet{2006ApJ...642...63Lee} suggest slightly lower halo
masses may be plausible ($M_{halo}\,\sim\,10^{11}$\msun) however the
uncertainties are too large to see a significant difference between
the \zth\ and \zfv\ populations. This conclusion is similar to that of
\cite{2004ApJ...611..685Ouchi} who found an increase (of marginal
significance) in the correlation length with increasing redshift which
is consistent with a constant halo mass of $\sim$10$^{12}$\msun\ over
z=3-to-5.  Again, however, the uncertainties, especially for the highest
redshift LBGs, were large.  Given this, and in the absence of
dynamical measurements of \zfv\ LBGs, we adopt the \zth\ mass range
\citep[$10^{11.2-11.8}$\msun,][]{2005ApJ...619..697Adelberger} for the
dark matter halos for \zfv\ LBGs.  Given that the comoving number
density of underlying dark matter halos of sufficient mass to host
such LBGs is evolving strongly from z\,=\,5 to 3
\citep{2002MNRAS.336..112Mo}, the bias of the LBGs must be increasing
with redshift \citep{2004ApJ...611..685Ouchi}.

The uncertainties in the clustering statistics of LBGs at
\zfv\ are still relatively large. 
The determination of accurate halo masses at \zfv\ will come from a
analysis of the clustering statistics of a sufficiently large sample
of either spectroscopically-confirmed or photometrically-robust LBGs
\citep[\rm{e.g.}][]{2004ApJ...611..685Ouchi}. As we have shown in this
paper, the latter requires a sample defined using both optical and infrared
bands to exclude low redshift galaxies. Until this is done
comprehensively, clustering statistics from samples selected by
optical photometry alone or those that only include shallow infrared
data, should be interpreted with caution.

\subsection{Evolution of the global star formation rate and stellar mass densities}

\subsubsection{Global star formation rate density}

\begin{figure}
\includegraphics*[width=8.5cm]{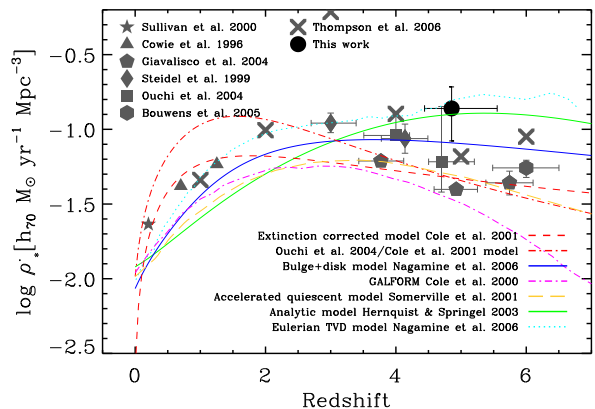}
\caption{Evolution of the star formation rate density with
  redshift. The SFRD determined from this study is shown (filled black
  circle, see text for details of its estimation) with respect to
  previous estimations and theoretical models. We emphasise the high
  redshift tail of the SFRD(z) and therefore only a handful of the
  numerous measurements at z$<$2 are shown. For consistency, all data
  and models are normalised to a Chabrier/Kroupa IMF. All data are
  correspond to an integration of the \zth\ LBG luminosity function
  \citet{1999ApJ...519....1Steidel} to 0.1L*, except for
  \citet{2004ApJ...600L.103Giavalisco} and
  \citet{2005astro.ph..9641Bouwens} which are to 0.2L* and
  0.04L$_{z=3}$*, respectively. Measurements from the literature were
  estimated using scaling relations to the UV luminosity and account
  for an extinction of E(B-V)=0.15, as per
  \citet{2006astro.ph..3257Nagamine}.}
\label{fig:sfrd}
\end{figure}

The evolution in the star formation rate density (SFRD) with redshift
has been a crucial tool in our understanding of the evolution of
galaxies in an ageing Universe. The ground-breaking 'Madau-plot'
\citep{1996MNRAS.283.1388Madau} and versions thereafter have clearly
defined the era of peak star formation activity in the Universe
\citep[z\,$\sim$\,1-2,
  e.g.][]{1997ApJ...486L..11Connolly,2004Natur.428..625Heavens},
however constraining the SFRD at higher redshifts is still an issue of
debate with direct observed estimates suggesting a decline for
z\,$>$\,3, while estimates based on integrating over the luminosity
function and theoretical predictions
\citep{2004ApJ...600L.135Somerville,2006astro.ph..3257Nagamine} imply
a constant value or shallow decline beyond the peak at z\,$\sim$\,2
(e.g. \citealp{1999ApJ...519....1Steidel,2004MNRAS.355..374Bunker,2004ApJ...611..660Ouchi,2004ApJ...600L.103Giavalisco,2005astro.ph..9641Bouwens,2006Natur.443..189Bouwens};
see also \citealp{2006ApJ...651..142Hopkins} for a compilation and
references therein). The model of \citet{2003MNRAS.341.1253Hernquist}
and the TVD hydrodynamical simulations of
\citet{2006astro.ph..3257Nagamine} suggest the SFRD might still be
increasing at \zfv.

Our results from the SED modelling of \zfv\ LBGs has an impact on this.
The SFRD derived from the best-fit star formation rates for LBGs in
our sample is compared in Figure \ref{fig:sfrd} to several recent
observational determinations and model predictions from the
literature.  This is determined from the total SFR for our robust
sample, corrected for the number of true \zfv\ LBGs we expect in our
total sample of 109 sources (all except the 40 per cent interloper fraction),
incompleteness at our magnitude limit (90 per cent\footnote{Gwyn, determined
  from the v1.0 release of the ACS data see
  http://www.astro.uvic.ca/grads/gwyn/virmos/cdfs/index5.html.}) and
accounting for galaxies to $\rm 0.1$\,L*\footnote{assuming the
  \zth\ luminosity function of \citet{1999ApJ...519....1Steidel} ($\alpha$=-1.6) since
  the faint end slope of the \zfv\ luminosity function remains
  uncertain.}, over our survey volume.  Nominally, the survey volume
is given by the area of the field, multiplied by the depth in redshift
space between the minimum and maximum redshifts that can satisfy our
initial selection criteria.  However, not all sources would remain in
our magnitude limited sample because of cosmological dimming and
increased intervening hydrogen opacity with redshift
\citep{2003ApJ...593..630Lehnert}.  Making the assumption that all
sources are at the maximum possible redshift to remain in the sample
(\rm{i.e.} they are close to the magnitude limit of $i_{AB}=26.3$) gives a
volume probed of $\sim$40 per cent the maximum notional volume
\citep[consistent with fractions determined from detailed
  calculations][and E. Stanway, private
  communication]{2003ApJ...593..630Lehnert}.  The star formation rate
density accounting for this effective survey volume is shown if Figure
\ref{fig:sfrd}. For consistency with the models, the SFRs determined
have been corrected to a Chabrier/Kroupa IMF, rather than the Salpeter
IMF we have used in the modelling, by dividing by a factor of 1.6
\citep{2006astro.ph..3257Nagamine}.
The high SFRD determined supports no more than shallow
decline in the SFRD at high redshift, consistent with the theoretical
predictions of \citet{2006astro.ph..3257Nagamine} and
\citealp{2003MNRAS.341.1253Hernquist}, however lies above the
semi-analytic accelerated quiescent model of
\citet{2001MNRAS.320..504Somerville}.

Star formation rates can also be derived by scaling the intrinsic
rest-frame UV luminosities of sources. The additional data points
shown in Figure \ref{fig:sfrd} are derived as such.
For starbursts
younger than $\rm \sim 100~Myr$, the ratio of star formation rate to
UV luminosity declines with age, levelling off at $\rm \sim 100~Myr$.
This is due to the time it takes to build up a UV luminous stellar
population in such a galaxy, the levelling-off occurring once their
birth and death rates match.  In part owing to this effect, a
\zfv\ sample matched in intrinsic UV luminosity with a \zth\ sample
will have higher star formation rates for the typically younger ages
we infer.  Here we note that the UV luminosity density declines by a
factor of $\sim 3$ between $z = 4$ and $z = 5$
\citep{2003ApJ...593..630Lehnert}.  Because of the younger ages at
higher redshifts, this must be an upper limit to the decline in SFRD,
at least for the more luminous LBGs.  A detailed exploration of this
issue is left to a forthcoming paper (N.M. F\"orster Schreiber et al.,
in preparation, Paper II).

\subsubsection{Global stellar mass density}

\begin{figure}
\includegraphics*[width=8.5cm]{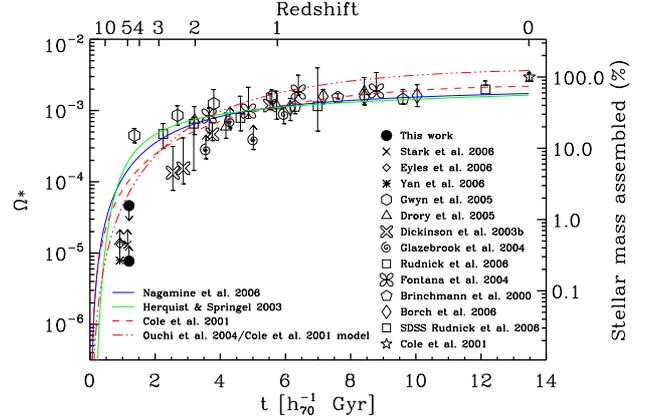}
\caption{Evolution of the stellar mass density with redshift. The left
  y-axis is plotted in the dimensionless $\Omega$*=$\rho$*/$\rho_c$*
  parameter and the right shows the percentage of stellar mass that
  has been assembled in the Universe with respect to the z=0 estimate
  from \citet{2001MNRAS.326..255Cole} (open star). All data and models
  correspond to a Chabrier/Kroupa IMF. The upper and lower limits
  determined from this study are plotted as (filled circles). The
  models are the integral of the SFRD(z) which are modified to account
  for mass loss from evolved stars.}
\label{fig:mstard}
\end{figure}

The evolution in the global stellar mass density with redshift places
strong constraints on models of the stellar mass assembly in the
Universe \citep{2003ApJ...587...25Dickinson}. Approximately 50 per cent of
the stellar mass in galaxies had formed by z$\,\sim\,$1-1.5, with
$\sim$10 per cent in place by
\zth\
\citep{2003ApJ...587...25Dickinson,2003ApJ...599..847Rudnick,2006ApJ...650..624Rudnick}.
Accurately constraining the evolution of this property with redshift
requires mass-selected (or limited) samples at different epochs. While
our sample of luminous LBGs is not mass selected, we can nevertheless
place limits on the formation of stellar mass soon after the end of
reionisation. We have determined the stellar mass in our survey volume
after applying the incompleteness corrections described in the
preceding section (with the exception of the correction to 0.1L$^{\star}$). The
stellar mass density at \zfv\ is
$\rho*\sim\,1.5\,\times\,10^{6}$\,\msun\,Mpc$^{-3}$
($\Omega*\,\sim\,1.1\times10^{-5}$), for a Salpeter IMF.  For
consistency with the published modelling, \rm{e.g.}
\citet{2006astro.ph..3257Nagamine}, we reduce this value by 1.4 to
account for the difference between the Salpeter IMF and the
Chabrier/Kroupa IMF used in the models, as shown in Figure
\ref{fig:mstard}. This limit represents $\sim$0.3 per cent of the
total stellar mass of galaxies seen today and it is consistent with
recent estimations at high redshift
\citet{2006astro.ph..4250Stark,2006astro.ph..7306Eyles,2006astro.ph..4554Yan}.

The stellar mass density derived is a lower limit partly because of
the stochasticity discussed in Section \ref{dutycycle}. The relatively
short duty cycle of these sources suggests that up to $\sim$12-in-13
LBGs at \zfv\ may be missed in our survey. Accounting for this
incompleteness places an upper limit on stellar mass density at this
epoch for our sources (see Figure \ref{fig:mstard}). Even after
accounting for this, our derived upper limit falls below the
predictions of the evolution of mass density with redshift obtained
from the integration of the models that plausibly explain the SFRD(z)
to high redshift (see Figures \ref{fig:mstard}). This is to be
expected given our sample is not selected to be mass complete.
Extending the mass function to lower masses would again increase the
mass density but at a lower level (a factor of a few) compared to the
duty cycle correction. Moreover, at z$\sim$2 as much of the stellar
mass density contained within UV-selected samples
\citep{2005ApJ...626..698Shapley} lies in the population of massive
distant red galaxies \citep{2006ApJ...650..624Rudnick}, which are
largely missed by UV selection techniques. Comparably massive galaxies
($>$10$^{11}$\msun) in the UV selected sample of
\citet{2005ApJ...626..698Shapley} are an order of magnitude less
numerous than the DRGs. Whether such massive galaxies dominate the
mass density at \zfv\ remains to be established.

The apparent discrepancy between these theoretical predictions and the
observed estimates for z\,$>$\,1 has been attributed to incompleteness
of the observational data
\citep[\rm{e.g.}][]{2006astro.ph..3257Nagamine}, or deficiencies in
the models, predicting too much stellar mass at high redshift, despite
reproducing the SFRD(z). \citet{2006ApJ...650..624Rudnick}
comprehensively discuss both observational estimates and model
predictions highlighting the potential causes for the discrepancies
from both perspectives \citep[see also the discussion in
][]{2005AJ....130.1337Gwyn}.  Notwithstanding, based on the
observational data alone, our upper limit implies that $\sim$1.5 per
cent of the present-day stellar mass density could have already been
been assembled in luminous LBGs in the first 1.2\gyr, rising by an
order of magnitude $\sim$1\gyr\ later
\citep{2006ApJ...650..624Rudnick}. This suggests that once the first
galaxies had formed, star formation and stellar mass assembly
proceeded at a prodigious rate in the early Universe.

\subsection{Mass and luminosity densities}
\label{size}

We measured the sizes of LBGs in the robust sample using the ACS
$z$-band images. Using the spectroscopic and photometric redshifts,
the mean (proper) size of the \zfv\ LBGs is found to be 1\,kpc.  This
is a 40 per cent decrease from the mean size of LBGs at redshift 3
($<r_e>\sim$1.7\,kpc), which is fully consistent with the known size
evolution of UV-selected galaxies \citep{2004ApJ...600L.107Ferguson,
2004ApJ...611L...1Bouwens, 2004MNRAS.347L...7Bremer}.  

\cite{2004ApJ...604..521Trujillo} found that there was no or only weak
evolution of the stellar mass densities (mass/r$_e^3$) of galaxies with
masses $>$10$^{10}$\msun over the redshift range of 0 to $\sim$3.  While
the results of \cite{2004ApJ...604..521Trujillo} are incomplete for lower
mass galaxies, we that the stellar mass densities for our LBGs at z=5 are
comparable to that found by \cite{2004ApJ...604..521Trujillo}.  However,
their UV luminosity densities (luminosity/r$_e^3$) are significantly
higher than LBGs at \zth\ with \zfv\ LBGs producing as much luminosity
as \zth\ LBGs in one-fifth of the volume. This difference cannot be
attributed to a low surface brightness, extended component since the
ACS images are significantly deeper than our magnitude limit for
selection \citep[see also][]{2004ApJ...611L...1Bouwens}.

The fact that the stellar mass densities are comparable to
star-forming galaxies at low redshifts, while the observed luminosity
densities are significantly higher, is indicative that the
mass-to-light ratio of such galaxies is evolving with redshift
\citep{2004ApJ...604..521Trujillo}. LBGs (and DRGs) at \zopen2-to-3 have
$<$M/L$_V$$>$ ratio of $\sim$0.5\,-\,1
\citep{2004ApJ...604..521Trujillo,2005ApJ...626..698Shapley} while we
find by \zfv\ LBGs have $<$M/L$_V$$>$ ratio of 0.1. In the
hierarchical framework, stellar mass assembly is likely to have
occurred in episodic bursts. Low mass-to-light ratios, such as those
we are finding for the \zfv\ LBGs, are expected for galaxies at high
redshift which are forming stars in a recent burst.  These findings
are consistent with our contention that these are young (and likely
without a significant intermediate age population, see Appendix
\ref{hidden}), unextincted intensely star-forming galaxies.

\subsection{Lyman alpha emitting galaxies}

The spectra of 9 of 15 LBGs with spectroscopic redshifts show an
emission feature with an asymmetric line profile and continuum break
consistent with the being Ly$\alpha$.
For \zfv\,-\,6 LBGs, \citet{2006ApJ...645L...9Ando} find that luminous
LBGs are Ly$\alpha$ deficient, with the separation between those which
are Ly$\alpha$ emitters occurring at $\ga$M*
($M_{AB, 1400\AA} \sim $-21.5 to -21.0~mag).
They postulate that more luminous LBGs are more chemically evolved
systems where dust absorbs the Ly$\alpha$ emission, although
alternative explanations cannot be ruled out.  This is consistent with
the ideas put forward by \citet{2006Natur.440..644Mori} who suggest
that Ly$\alpha$ emitting and Lyman-break galaxies are evolutionary
stages in the chemical enrichment of galaxies that trace the evolution
of primeval irregulars to present day ellipticals. From N-body
hydrodynamical simulations with stellar population synthesis they
state that the Ly$\alpha$ is produced by cooling shocks in the early,
sub-solar metallicity evolutionary phase for ages $<$\,300\myr. Then,
up to 1\gyr, the stellar continuum increases and the metallicity
becomes solar.

We would therefore expect to see a trend in our derived ages and the
equivalent width of Ly$\alpha$ in our sample with older galaxies
having lower equivalent widths. Using the ESO-FORS2 spectroscopy, we
have crudely estimated the Ly$\alpha$ equivalent widths (ignoring
underestimation caused by the blue-shifted absorption due to
intervening opacity along the line of sight) of the Ly$\alpha$
emission to be -20\,$\le$\,EW(Ly$\alpha$)\,$\le$\,0\,\AA\, for all LBGs
in the robust sample, irrespective of absolute magnitude
(-24.0$<$$M_{AB, 1400\AA}$$<$-20.5). However, as spectroscopy has only
been obtained for the brightest LBGs, it is not surprising that we do
not see the extremely faint Ly$\alpha$ emitting galaxies with high
equivalent widths ($\la$-40\AA) seen for $M_{AB, 1400\AA} \ga $-21.5. 
There is no significant correlation between the strength of Ly$\alpha$
equivalent width with the nominal best-fit ages.
We do not find any strong difference in the distributions of mass or
star formation rate suggesting these properties do not influence the
presence or absence of Ly$\alpha$. Furthermore, we find no evidence
that the 'break-only' galaxies have a higher average extinction than the galaxies
exhibiting Ly$\alpha$ in their spectra and therefore attenuation by
dust may not be the sole reason why there is a lack of Ly$\alpha$
emission in the break-only LBGs. Rather local conditions must
influence the presence and strength of Ly$\alpha$ emission. It may
well be that young galaxies have a high covering fraction of neutral
gas which becomes successively more ionised as the galaxy ages. The
effect of winds and the geometry of the gas may also play a crucial
role in the range of ages seen in galaxies without significant
Ly$\alpha$ emission.

\section{Discussion}
\label{discussion}

We present detailed analysis of the properties of LBGs at \zfv\ with
robust and reliable photometry spanning the rest-frame UV to
optical. Through a probability analysis we are able to constrain the
properties of the ensemble and find that 'young' ($<$\,100\myr) and
moderately massive ($\sim 2\,\times\,10^{9}$\msun) galaxies are in
fact more than twice as prevalent at \zfv\ (70 per cent) than at
\zth\ (30 per cent). This increased fraction of young and moderately massive
galaxies suggests that at redshift 5 we are seeing an era of
widespread stellar mass assembly in the early evolution of
galaxies. Based on these properties, we discuss three key
characteristics of this population that emerge from our analysis.

\subsection{Progenitors of present day early type galaxies or bulges}

The luminous \zfv\ LBGs studied here are undergoing a period of
intense, recent star formation that dominates their rest-frame
UV-to-visible emission. Their compact sizes, as measured from the HST
images, show that the UV emitting regions are small (mean and median
half-light radius of $\sim$1kpc). In their short lifetimes, each of
these galaxies has already assembled $\ga$\,2\,$\times$10$^9$\msun\ of
stars within this radius. The inferred stellar mass surface density
($\mu_{\star}\,\sim\,$6$\,\times$10$^{8}$\msun\,kpc$^{-2}$) is
comparable to the bulge and spheroidal components seen in present-day
typical (L*) galaxies \citep[\rm{e.g.}
  $<\mu_{\star}>\sim$5$\times$10$^{8}$\msun\,kpc$^{-2}$,][]{2002AJ....124.1360Galaz}.

From a study of nearly 400,000 low redshift galaxies drawn from the
Sloan Digital Sky Survey (SDSS),
\citeauthor{2006MNRAS.367.1394Kauffmann} have shown that significant
differences between the structural parameters of early and late type
galaxies
are highly dependent on their mass surface densities, rather than the
total mass of the galaxies. Specifically, they find that the
concentration parameter ($C>2.5$) and stellar mass surface densities
($\mu_{\star}>3\times 10^{8}$ M$_{\sun}$ kpc$^{-2}$) correspond to the
regime of galaxy spheroids and bulges, independent of the total
stellar mass of these systems.  Above this threshold of mass surface
density, they find that star formation is increasingly suppressed in
local galaxies (the specific star formation is low) and must have
ceased many Gyrs ago. As a consequence,
\cite{2006MNRAS.367.1394Kauffmann} hypothesize that with increasing
compactness and surface density of the galaxy, stars were formed in
short, vigorous episodes at high redshift, with extended periods of
inactivity \citep[][characterized this as the gas consumption time
  scale, t$_{cons}\propto
  \mu_{\star}^{-1}$]{2006MNRAS.367.1394Kauffmann}.  This analysis is
particularly appropriate here since we do not find high stellar masses
which is often taken as evidence for being progenitors of early type
galaxies; we instead find very high mass surface densities.  Given the
circumstantial evidence that the \zfv\ LBGs in our sample drive winds
(see Section \ref{winds}), have young ages and short duty cycles, it
is also plausible that these galaxies will grow in strong bursts of
star-formation. So while we cannot know the evolutionary path of this
population of galaxies, their overall mass surface densities suggest
that they are the progenitors of early type galaxies or bulges, and
favours the ``inside-out'' galaxy formation scenario.

The star formation that dominates the UV-to-visible SEDs of the
\zfv\ LBGs in our sample is likely to be such a burst of star
formation in which a significant fraction of the stellar mass of the
system has been formed. A simple estimate of the relating the
dynamical time-scale to the mass and size of a galaxy is

\begin{eqnarray}
t_{dynamical}\,=\,\frac{r}{V}\,\approx\,\frac{\eta^{1/2}\,r^{3/2}}{M^{1/2}\,G^{1/2}}, 
\end{eqnarray}\label{eqn:tdyn}

where $V$ is the characteristic (virial) velocity of the system,
$\eta$ is a geometrical correction factor that depends on the mass
distribution within the system, $r$ is the radius or physical scale of
the system, $M$ is the mass of the system, and $G$ is the
gravitational constant.  Depending upon the mass distribution for a
given galaxy, $\eta$ is $\approx$ 3-5,
\citep{1987gady.book.....Binney,1997MNRAS.285..779Rix}.  With a median
measured half-light radius of $\sim$\,1\,kpc and a mass of
2\,$\times$\,10$^9$\msun, we estimate that the dynamical time-scale of
the galaxies in our sample is of order 20\myr.  
Therefore, the intense star formation in
the majority of the \zfv\ LBGs (age$\sim$100\myr) has been typically
proceeding for approximately one to $\sim$5 dynamical
time-scales. Starbursts, the youngest, large-scale star
star forming events seen in nearby galaxies have comparable durations
\citep{1996ApJ...472..546Lehnert,1998ARA&A..36..189Kennicutt,2003A&A...399..833Foerster}.
A system that has assembled a significant fraction of its stellar mass
over the last few dynamical time-scales is in essence a galaxy in the
process of formation.  Thus, the young galaxies in
our sample are likely to be members of the long sought-after
population of primeval galaxies in the process of significant stellar
mass build up. 

Further support of the progenitor scenario arises from analysis of the
clustering of \zfv\ LBGs.  The clustering length of \zfv\ LBGs implies
that L$>$L* LBGs at \zfv\ reside in dark matter halos with masses of
order 10$^{11-12}\,h_{70}^{-1}$\msun\ \citep{2004ApJ...611..685Ouchi,
  2006ApJ...642...63Lee}. While this is comparable to the dark matter
mass halo of the Milky Way and to the halos of \zth\ LBGs
\citep{2005ApJ...619..697Adelberger}, this should not be interpreted
as supporting that LBGs are the precursors of galaxies like the Milky
Way or \zth\ populations. Linking a population of galaxies at one
redshift with another population at another is fraught uncertainty
\citep[see for example, ][]{2002ApJ...577....1Moustakas}.  If these
galaxies are the progenitors of galaxies like the Milky Way, this
would require an evolutionary path whereby the clustering amplitude
and halo mass does not grow, the bias decreases, but the baryonic mass
would have to increase by over an order of magnitude. In a galaxy
conserving model for the growth of structure
\citep{1996ApJ...461L..65Fry} and if we think of the estimated bias
of the LBG population as z$\sim$5 as its bias at birth, would suggest
that bias would decrease with decreasing redshift while the
correlation scale would increase \citep{1996ApJ...461L..65Fry,
  2002ApJ...577....1Moustakas}.  In such a model, the evolution of
both the bias and correlation scale suggest that the \zfv\ LBGs could
be the progenitors of the red galaxies at \zopen1 and massive early
type galaxies at z$\approx$0 \citep{2004ApJ...617..765Coil,2005ApJ...630....1Zehavi}.  The conclusions would be similar in
hierarchical merging models \citep{2004ApJ...611..685Ouchi,
  2002ApJ...577....1Moustakas}.  In either model, it is difficult to
say that the \zth\ and \zfv\ are directly related (except for the
large uncertainties in the clustering and bias estimates for the
\zfv\ LBGs).

In the hierarchical merging scenario, phase-space density arguments
suggest that massive, present-day elliptical galaxies have formed from
the merging of high redshift disk galaxies, which are smaller and
denser than disk galaxies seen today \citep{1998MNRAS.296..847Mao}.  If
the \zfv\ LBGs represent the early stages of the formation of present
day bulges and spheroids they should have core phase-space densities
that are consistent with their present-day descendants. A
significantly lower phase space density would rule out a progenitor
scenario. Following \citet{1986ApJ...310..593Carlberg} and
\citet{1998MNRAS.296..847Mao} the core phase-space density of the
typical \zfv\ LBG in our sample is $\sim$10$^{-6}$\msun\ (pc km
s$^{-1}$)$^{-3}$. This is consistent with the core phase-space
densities of massive, bright present day ellipticals and the central
region of the Milky Way
\citep{1986ApJ...310..593Carlberg,1998MNRAS.293..429Wyse,1998MNRAS.296..847Mao,2005MNRAS.361..997Avila}. While
we cannot constrain the evolutionary path of these LBGs, this
similarity between the core phase-space densities of our \zfv\ LBGs
and present day bulges and ellipticals, strongly suggests that the
central population in these systems is already in place at high
redshift.

The young \zfv\ LBGs may therefore represent the formation of the
central regions of present-day massive galaxies at the earliest stages
of their evolution.  If so, their abundance at a look back time of
12-13\gyr\ is consistent with the age of the dominant stellar
population in the Galactic Bulge \citep{2005NewAR..49..465Rich} and
its proposed formation in short bursts 1-2\gyr\ after the Big Bang
\citep{2006astro.ph..9052Zoccali}.

\subsection{Enrichment of the intra- and inter-galactic medium}

\subsubsection{Winds}
\label{winds}

These galaxies are young relative to the age of the Universe at the
epoch they are observed ($\sim$1.2\gyr\ at the median redshift).  The
young ages derived for the majority of \zfv\ LBGs imply that these
systems began appreciable star formation at z$_f\,\sim\,$6 with the
bulk of their stars formed at z\,$<$\,6. As such, if reionisation was
complete at \zfv.8
\citep{2001AJ....122.2850Becker, 2001AJ....122.2833Fan}, the intense
star formation episode studied here could only significantly
contribute to the end of reionisation and these galaxies are unlikely
to be the perpetrators. However, these young galaxies do have an
impact on the high redshift IGM.  These compact galaxies are
experiencing widespread and recent star formation at prodigious rates.
The UV emitting regions are small ($\sim$1~kpc) and consequently these
\zfv\ LBGs have high rest-frame UV surface brightnesses. Given the
typical star formation rates inferred from the SED modelling, their
star formation rate surface densities (star formation rate per unit
area) are several tens to hundreds\sfrarea. This is far higher
than observed star formation rate surface densities of local galaxies
that are known to be driving vigorous gaseous outflows
\citep[$>$\,0.1\sfrarea; ][]{2001ASPC..240..345Heckman}.  Thus,
young \zfv\ galaxies host strong outflows, expelling the metals
created by stellar nucleosynthesis into the surrounding medium. As
such, these vigorous starbursts must contribute to the early chemical
enrichment of the IGM which is supported by the detection of ions of
metals such as Carbon in the IGM at redshifts as high as $\sim$\,5
\citep[and references therein]{2005AJ....130.1996Songaila}.

\subsubsection{Galaxy-scale metal-mixing \& metal-free star formation}

The first stars in the Universe must have been born from
gas of primordial abundance (so-called population III
stars). Recent detections of the signatures of population III stars
in the composite spectrum of \zth\,-\,4 Lyman break galaxies suggest
that such a population comprises 10-30\% of the stellar mass, implying
that galaxy-wide mixing of metals is inefficient on time-scales of
a billion years \citep{2006Natur.440..501Jimenez}\footnote{Note:
our assumed 0.2Z$_{\odot}$ corresponds very well to the combined
Z$\sim$~0Z$_{\odot}$ and Z$\sim$0.4Z$_{\odot}$ mixed model of
\citet{2006Natur.440..501Jimenez}.}. Most models of galaxy formation have
assumed efficient intra-galactic mixing of metals and do not predict
metal-free star formation at redshifts significantly below \zfv\
\citep{2004ApJ...605..579Yoshida}.

While the star-formation intensities suggest that our young galaxies are
likely to drive outflows and pollute the IGM with metals, and presumably
their own interstellar media, their very young ages imply that this
material does not reach far beyond galactic scales.  We can quantify this
statement by making some simple assumptions about the outflows driven by
these high intensity star-forming galaxies in analogy with local starburst
galaxies.

The temperature of the outflowing gas can be estimated by simple energetic
arguments \citep[\rm{i.e.}, energy injected rate of the intense star formation
equals the rate at which material -- both ambient and stellar ejecta --
is thermalised;][]{2001ASPC..240..345Heckman, 1999ApJ...526..649Moran}

\begin{eqnarray}
T_{wind}\approx 0.4 m_H \frac{\dot{E}}{k\dot{M}} = 5.4 \times 10^7 \frac{\dot{E}_{43}}{\dot{M}_{10}} f_{loading}^{-1} \,{\rm K}
\end{eqnarray}\label{eqn:Twind}

where $m_H$ is the mass of hydrogen, $\dot{E}$
is the energy injection rate, and $\dot{M}$ is the mass outflow rate, $k$
is Boltzmann's constant, and  $f_{loading}$ is the fraction of hot gas
that results from ``mass loading'' due to sweeping up and entraining
ambient interstellar material. Using the models of \cite{1999ApJS..123....3Leitherer}, at the
median age and star-formation rates we have estimated form fits to the
SEDs of the robust sample of LBGs, the typical energy and mass
injection rates are about 10$^{43}$ ergs s$^{-1}$ (=$\dot{E}_{43}$)
and about 10\sfr (=$\dot{M}_{10}$), respectively to which we have scaled
equation \ref{eqn:Twind}.

Similarly, equating the kinetic energy of the gas
to the energy injection rate yields an estimate of the outflow velocity:

\begin{eqnarray}
v_{wind}= (2 \frac{\dot{E}}{\dot{M}})^{1/2} \sim 1800 f_{loading}^{-1/2} (\frac{\dot{E}_{43}}{\dot{M}_{10}})^{1/2} \,{\rm km}\,{\rm s}^{-1}
\end{eqnarray}\label{eqn:vwind}

The mass loading of local starburst winds has been estimated to be about
a factor of a few to 10 \citep{1999ApJ...526..649Moran}.  This would
decrease the wind velocity accordingly.  If we assume that winds are
loaded by about a factor of 5, then the wind velocity is $\approx$800
km s$^{-1}$.  In the typical age of the LBGs, the wind would traverse
$\sim$15~kpc.  Therefore, in principal, it is possible for the wind to
span and enrich the ambient interstellar medium with metals.  However,
we have assumed that the mass loading is a simple drain on the energy
of the wind material itself.  In reality, the ambient ISM will be an
inhomogeneous medium with clumps of higher density material.  The diffuse
ISM will get swept up while the clouds will be entrained and accelerated
by the ram pressure of the wind material itself.  In such an situation,
it is important to consider both how a continuous medium and clouds are
accelerated in the outflowing gas.  Following, \rm{e.g.}, \citet{1980pim..book.....Dyson},
expansion speed of the shell and energy content of the bubble are
related as:

\begin{equation}
v_{shell} \sim 240 \dot{E}_{43}^{1/5} \ n_{0}^{-1/5} \ t_{7}^{-2/5}
\ {\rm km} \ {\rm s}^{-1},
\end{equation}

where $\dot{E}_{43}$ is the (constant) energy injection rate in units
of 10$^{43}$ ergs s$^{-1}$, $n_0$=1.0 cm$^{-3}$ is the ambient ISM
density in cm$^{-3}$ and $t_{7}$ is the injection time in units of
10\myr.  At such velocities, swept-up ambient ISM will traverse about
5~kpc over the age of the typical LBG in our sample.  Clouds
accelerated in the wind will be dispersed over a similar size scale.
Since this is only a factor of a few larger than the half light radii
of the galaxies in our sample, given the strong collimation of
superwinds observed in the local Universe
\citep[\rm{e.g.},][]{1990ApJS...74..833Heckman, 1996ApJ...472..546Lehnert},
it is unlikely that the metals in the outflow will be mixed
significantly with the infalling or ambient gas.

While even the swept-up ambient ISM is unlikely to be efficiently
mixed by the pressure and outflow of the intense star-formation,
another interesting time scale to investigate is the cooling time of
the hot superwind gas.  If the gas were to have a long cooling time,
then it would be unavailable for star-formation.  This would effectively
sequester the metal enriched gas from the stellar mass loss from being
incorporated into star-formation over a cooling time scale.  We estimated
previously that the wind material itself, modulo the mass loading, will
have a temperature of $\approx$\,5\,$\times$\,10$^7$\,K.  This plasma will also
be very rare.  We can crudely estimate the density of the gas using
the relationship from \citet{1985Natur.317...44Chevalier}, namely,

\begin{eqnarray}
n_{e,wind} &\approx& \dot{M}^{1.5} \dot{E}^{-0.5} R^{-2} \\
&\approx& 6 \times 10^{-2} \dot{M}_{10}^{1.5} \dot{E}_{43}^{-0.5} R^{-2} {\rm cm}^{-3}
\end{eqnarray}\label{eqn:density}

where we have scaled the radius to 1\,kpc, $ R_{1 kpc}$, approximately
the half-light radius of the typical LBG in our sample and the area over
which energy is being injected by massive stars. If the wind that has
reached large distances from the LBG is mass loaded, the density will be
of course higher (likely be a factor of a few to 30).  The various ranges
of density, depending on radius and mass loading is of-order 10$^{-2}$\,to\,
10$^{-4}$, similar to what has been estimated in local starburst galaxies.
Using this crude estimate, we can investigate the likely cooling time
of the wind using standard cooling arguments.
For an equilibrium plasma at a
temperature of 10$^{6.7}$\,to\,10$^{7.7}$K and density of 10$^{-2}$\,to\,10$^{-4}$,
the cooling time is, roughly 10$^8$ years to a few\,$\times$\,10$^9$
years.  Over that duration, the star-formation will be able to build
up about 10-30\% of the mass of LBGs at \zth\ and do so with
much of the metals generated by the intense star-formation at higher
redshifts locked in hot gas that takes about 1\gyr\ to cool.  Of
course, the velocities estimated are also to likely be approximately
or greater than the escape velocities of the dark matter halos and
much of the enriched material may escape.  These tight timing
constraints are consistent with evidence of both a significant
fraction of population III stars in lower redshift Lyman break
galaxies and inefficient metal-mixing on intra- and inter-galactic
scales \citep{2006Natur.440..501Jimenez}.

\section{Summary}

We present the properties of a robust sample of LBGs at \zfv\ selected
to $i_{AB}\,<\,26.3$ in the Chandra Deep Field South. On average, LBGs
at \zfv\ are $\sim$10 times less massive ($\sim$10$^9$\msun) and are
significantly younger ($<$100\myr) than similarly luminous LBGs at
\zth. While LBGs with such low masses and young ages are not unknown
at \zth\ they are far less common, such systems comprise $\ga$70\% of
LBGs at \zfv\ and in contrast only 30\% at \zth. Their short duty
cycles suggest that the \zfv\ population must be highly stochastic and
that samples of \zfv\ LBGs may be highly incomplete, with only
$\sim$1-in-13 LBGs being detected. Considering this implies that up to
$\sim$1.5\% of the present-day stellar mass density can be accounted
for by luminous LBGs \citep[\rm{c.f.} 15\% a
  \gyr\ later][]{2006ApJ...650..624Rudnick}. The abundant fraction of
young and moderately massive \zfv\ LBGs are likely to be experiencing
their first (few) generations of large-scale star formation and are
accumulating their first significant stellar mass. Their formation
redshifts (z$\sim$6-to-7) suggests these galaxies could have contributed
only to the end of reionisation. They are extremely compact as
measured from the UV ACS images ($r_{1/2}\,\sim\,$1~kpc) giving rise
to high stellar mass and star formation rate surface densities. The
stellar mass surface densities and core phase-space densities are
comparable to the spheroidal components of present day L* galaxies and
suggests these systems are the progenitors of early-type galaxies or
bulges, favouring the inside-out galaxy formation scenario. The high
star formation rate surface densities implies these sources are
driving winds that enrich the inter and intra-galactic media with
metals. Their young ages are consistent with inefficient metal-mixing
on galaxy-wide scales. Therefore it is plausible that these galaxies
contain a significant fraction of metal-free stars as has been
proposed for \zth\ LBGs.

\section*{Acknowledgments}

We would like to thank the anonymous referee for their constructive
and insightful comments that substantially improved this paper.  We
thank Matthias Tecza and Elizabeth Stanway for their comments and
useful discussion.  Some of the data presented in this paper were
obtained from the Multimission Archive at the Space Telescope Science
Institute (MAST).  STScI is operated by the Association of
Universities for Research in Astronomy, Inc., under NASA contract
NAS5-26555. This work is based [in part] on observations made with the
Spitzer Space Telescope, which is operated by the Jet Propulsion
Laboratory, California Institute of Technology under a contract with
NASA. We acknowledge the ESO/GOODS EIS and spectroscopy projects
(13,17) which have been carried out using the Very Large Telescope at
the ESO Paranal Observatory under Program ID: LP170.A-0788 \&
LP168.A-0485.

\appendix
\section{Alternative modelling assumptions}
\label{alt}

To assess the sensitivity of our characteristic ensemble properties to
the model assumptions, we fit the SEDs of the \zfv\ LBGs with
different sets of input parameters, varying the star formation history
(SFH), extinction law and metallicity. As we are interested in the
ensemble properties, we focus on the effects of variations on the
median properties of the robust sample of \zfv\ LBGs. Degeneracies
between the model parameters make it difficult to discriminate between
SFHs, extinction laws or metallicities based solely on the reduced
chi-squared values ($\chi^{2}_{n}$). One must also keep in
mind that there are uncertainties in the stellar evolution, spectral
libraries, and dust properties themselves, especially important at
sub-Magellanic metallicities \citep[see the discussion
by][]{2003MNRAS.344.1000Bruzual} as well as regarding their validity
at \zfv.  Differences in goodness-of-fit, even systematic, may
reflect in part template mismatches.

We discuss the effects of varying the assumed parameters in the
following subsections and Table \ref{tab:models} summarises some of the key
variations. However, in view of all the uncertainties involved, the
numbers quoted in the following should not be taken at face value but
rather as indications of overall trends.  The interloper fraction (low
$z$ galaxies as well as star/QSO candidates) for the initial sample of
109 candidate \zfv\ LBGs is roughly 40\% in all cases; the median
$z_{\rm ph}$ for the robust sample is always around 4.8 except for
dust-free models with star formation time-scales longer than $\rm \sim
50~Myr$, for which it increases by about 0.2.  Only models without
dust extinction could affect our conclusion of young ages and could
make the ensemble as old or older than $\rm \sim 100~Myr$.  However,
such models are also those that lead to the poorest fits compared to
any other set of input parameters, although they cannot be ruled out
from a purely statistical $\chi^{2}$ argument. Typical stellar masses
for the ensemble of a few\,$\times \rm 10^{9}~M_{\odot}$ are derived
for all but the lowest metallicities explored, $Z = 0.002 - 0.02~{\rm
  Z_{\odot}}$.

\subsection{Star Formation History}

In addition to our adopted constant star formation rate (CSF) scenario,
we considered a suite of exponentially declining star formation rates
(SFRs) with $e$-folding time-scales ($\tau$) of 10, 30, 50, 100, 300, 500~Myr
and 1~Gyr, as well as the case of a single stellar population (SSP) formed
in an instantaneous burst.

Adopting the 0.2\zsun\ models and the SMC extinction law, the median
best fit age is a few tens of Myr, for all SFHs except for the SSP
where it decreases by a factor of $\approx$2.  This behaviour is not
surprising given the degeneracies between age and SFH: the best-fit
age gives a measure of the time elapsed since the bulk of stars that
dominate the observed SED was formed, so the derived ages tend to be
younger for models with shorter star formation time-scales.  Since
most of the LBGs in the robust sample have ages of a few tens of Myr,
any SFH with time-scale of that order or longer is effectively
equivalent to a CSF model and the derived ages are similar.  The
best-fit extinction changes little, with median $A_{V}$ values of a
few tenths of magnitude for $\rm \tau \ga 30~Myr$ down to 0~mag for
shorter time-scales.  The median SFR decreases by a very modest
$\approx 10\%$ between CSF and $\rm \tau = 30~Myr$ models, and drop by
a factor of 2 for $\rm \tau = 10~Myr$ (by definition, there is no star
formation for a SSP with non-zero age).  The median stellar mass shows
no systematic variations and remains roughly constant with star
formation time-scale, the largest difference amounting to $\approx
17\%$, smaller than the dispersion among individual objects in the
sample.

The trends of lower median best-fit ages, $A_{V}$ values, and SFRs
with star formation time-scales are qualitatively the same irrespective
of the metallicity or extinction law.  The exception is the median age
for dust-free $0.002~Z_{\odot}$ models, which increases from 3~Myr for
all SFHs to 6~Myr for an SSP; this difference is however at the limit
of the age sampling in our modelling and is not very significant.
The median SFRs drop by $\approx 15\%$ up to a factor of $\approx 5$
from CSF to $\rm \tau = 10~Myr$ models, depending on the metallicity
and extinction law.
For the stellar masses, the median values vary by a factor of 1.5 or
less between different SFHs for any given metallicity and extinction law
considered, and up to a factor of 1.8 for models without dust extinction.

\subsection{Extinction Law}
\label{extlaw}

Our choice of an SMC-type extinction law was motivated by
  consistency with the adopted metallicity of $Z = 0.2~{\rm
    Z_{\odot}}$.  Assuming the starburst extinction law of
  \citet[hereafter Calzetti law]{2000ApJ...533..682Calzetti} instead, which is widely applied
  in high redshift studies, the median best-fit $A_{V}$ values
  increase by a large amount up to about 1.2~mag (0.6~mag for an SSP)
  and at the same time, extremely young median ages of only a few Myr
  are derived.  The Calzetti law is much greyer (\rm{i.e.}, shallower) in
  the rest-frame UV than the SMC law, so that higher levels of
  extinction are allowed but this then requires younger ages to
  fit the very blue SEDs of our LBG sample.  As a consequence of
  the younger ages and higher $A_{V}$ values, both the median stellar
  mass and SFR increase, by factors of $2 - 3$ and $30 - 60$,
  respectively.

The properties and geometry of interstellar dust at $z \sim 5$ is
currently an open issue, and the application of extinction laws derived
for low redshift galaxies to high redshift systems is questionable.
Low to moderate extinction would be expected for young systems in which
it would be unlikely that significant amounts of widespread dust have
had sufficient time to form.  Moreover, the detection of Ly$\alpha$ in
the spectrum of some of these systems argues against strong attenuation
by dust.  There is empirical evidence that dust in $z \sim 2 - 4$ LBGs
exhibits SMC-like characteristics \citep{2003ApJ...587..533Vijh}.
At higher $z$ and given the young ages inferred for our LBG sample, it may
be that dust from Type II supernovae dominates the extinction, as proposed
for z$\,\sim\,$6 quasars by \citet{2004Natur.431..533Maiolino}.  However,
in the absence of determinations valid for star formation dominated
$z \sim 5$ LBGs, the SMC extinction law is a reasonable assumption.
It also leads to modest extinction, older ages, and lower SFRs that are
more plausible than the extremely young ages of a few Myr, and the high
$A_{V} \sim 1~{\rm mag}$ and SFRs $\rm \sim 1000~M_{\odot}\,yr^{-1}$
obtained with the Calzetti extinction law.

Models where we allowed extinction to be a free parameter fit the data
better than models without attenuation by dust, even though the
required best-fit extinction is modest for our adopted metallicity and
SMC extinction law ($A_{V} \sim 0.3~{\rm mag}$).  This is reflected in
the overall systematically higher $\chi^{2}_{n}$ values for the
zero-extinction case. Strictly speaking, the differences are not
statistically significant so that models without dust extinction
cannot be conclusively excluded. However, given that there is
spectroscopic evidence that the metallicity of LBGs is
0.2\zsun\ \citep{2004ApJ...610..635Ando}, a dust-free solution is
highly unlikely.

\subsection{Metallicity}

We explored the effects of metallicity with models for
$Z = 0.002, 0.02, 0.2, 0.4, 1~{\rm Z_{\odot}}$ (\rm{i.e.} those available
for the Bruzual \& Charlot models based on the Padova~1994 evolutionary
tracks and BaSeL 3.1 spectral library).  For this comparison, we chose
the Calzetti extinction law for all metallicities.  For any of the SFHs we considered,
the best-fit ages are slightly older at higher metallicities.  The highest
median extinction ($A_{V} \sim 1~{\rm mag}$) is derived for $Z = 0.2$ and
$\rm 0.4~Z_{\odot}$ models.  The median stellar masses decrease by a factor
of $4 - 5$ between $\rm 0.002~Z_{\odot}$ and $\rm Z_{\odot}$ while the
median SFRs vary by a factor of $3 - 8$ and are typically lower at both
ends of the metallicity range explored.  The overall trend for the median
stellar masses is preserved for models without dust extinction, but those
for the ages and SFRs are reversed.

The typical stellar masses of our robust sample LBGs are thus most 
affected by changes in the metallicity, but by only a
factor of 5 or less.  The median masses for our adopted
$\rm 0.2~Z_{\odot}$ metallicity, supported by spectroscopic results
of \citet{2004ApJ...610..635Ando}, lies at the middle of the range
inferred between $\rm 0.002~Z_{\odot}$ and $\rm Z_{\odot}$ for any
given SFH.

\begin{table*}
\begin{minipage}{\textwidth} 
\caption{The effects on the derived properties of SFR, mass, age and extinction of our \zfv\ LBGs with a selection of varying input parameters are summarised in this table.}
 \label{tab:models}
 \begin{tabular}{lccccc}
  \hline
Parameter & Variation & SFR & Mass & Age & A$_v$\\
\hline
Star Formation History \footnote{for 0.2\zsun\ templates and the SMC extinction law} & CSF &  few$\times$10 \sfr & 2$\times$10$^9$\msun & few$\times$10 Myr & few$\times$10$^{-1}$ mag \\
 & exp. declining $\tau >$30\myr & $\sim$few$\times$ 10 \sfr  & 2$\times$10$^9$\msun  & few$\times$10 Myr &  few$\times$10$^{-1}$ mag \\
 & exp. declining $\tau <$30\myr &  10 \sfr & 2$\times$10$^9$\msun & few$\times$10 Myr & 0 mag \\
 & SSP & - & 2$\times 10^9$\msun & 10 Myr & 0 mag \\
\hline
Extinction law & SMC &  few$\times$10 \sfr & 2$\times$10$^9$\msun & few$\times$10 Myr &  few$\times$10$^{-1}$ mag  \\
 & Calzetti extinction law & few-several$\times$10$^{2-3}$ \sfr & 5$\times 10^9$\msun & few Myr & 1-2 mag \\
\hline
Metallicity & 1\zsun              & few$\times$10 \sfr & 5$\times$10$^9$\msun & 200 \myr & 1-2 mag\\
            & 0.2 or 0.4 \zsun    & few$\times$10$^2$ \sfr & 2$\times$10$^9$\msun & 100 \myr & 1 mag \\
            & 0.002 or 0.02 \zsun & few$\times$10 \sfr  & 1$\times$10$^9$\msun & few tens \myr &  few$\times$10$^{-1}$ mag  \\
\hline
 \end{tabular}
\end{minipage}
\end{table*}

\section{Limits on the contribution of an old population}
\label{hidden}

The SEDs of all our \zfv\ LBGs are very blue in the rest-frame UV,
indicating the presence of on-going or very recent star formation.
The best-fit stellar masses (like all other derived properties)
represent those of the young and luminous populations that dominate
the integrated rest-frame UV to NIR light.  By selection, we would not
expect to have picked up old passively evolving galaxies. Therefore it
is unlikely that these luminous LBGs contain a significant, older,
less luminous component that significantly contributes to the stellar
mass. Moreover, the \zfv\ LBGs in our sample are extremely compact in
the GOODS-ACS data ($r_{1/2}\sim$1\,kpc), and given that they are
selected at an $i$-band magnitude significantly brighter than the
depth of these images, any old stellar population would have to follow
an extremely broad distribution to remain undetected. If any
significant old population is distributed over the same
extent as the UV luminous galaxy, then this would imply an extremely
high mass surface density ($\ga 10^{10}$\msun~kpc$^{-2}$), in place
already at \zfv. Such high mass surface densities are known only in
the inner regions ($\la$kpc-scale) of massive cD galaxies, and they
are not for typical starburst galaxies or \zth\ LBGs. This unlikely
high surface density strongly argues against a significant fraction of old
stars contributing to the galaxy mass, and suggests that the stellar
mass derived from the stellar synthesis modelling is representative of
the total mass of the galaxy.

Nevertheless, we can place extreme upper limits on the mass
contribution of such an old underlying population following a similar
procedure to that used for LBGs at
\zth\
\citep{2001ApJ...559..620Papovich,2001ApJ...562...95Shapley,2005ApJ...626..698Shapley}.
This consistent approach allows us to test whether such an old
contribution can explain the mass difference we have found which has
been discussed in Section \ref{comp}. For this we consider an extreme
'worst case' scenario, requiring the coincidence of a SSP which,
compared to scenarios of extended star formation, maximizes the
mass-to-light ratio of a ``hidden'' component for a given age.  For
this we assume the true photometry of the source is in fact at the
brightest 1$\sigma$ limit of our photometric uncertainties. These
uncertainties already include a generous 10\% to account for flux
calibration errors. Therefore, the limit that results is extreme and
while there is a chance that it can apply to an individual source, it
is highly unlikely to be valid for the population as a whole.

We consider the case of a passively evolving component with an age of
500\myr, the median for the one-third of our robust sample older than
100\myr, and comparable to the oldest ages for \zfv\ LBGs reported by
other authors as well
\citep[\rm{e.g.}][]{2005MNRAS.364..443Eyles,2006NewAR..50..127Yan}.  The
SED of a 500\myr\ old SSP peaks in the rest-frame optical, in
particular just red-wards of the Balmer/4000\,\AA\ break, and so would
contribute most to the observed fluxes in the $K_s$, $\rm 3.6~\mu m$,
and $\rm 4.5~\mu m$ bands.  We thus determined the upper limit for an
old component by adding to the nominal best-fit SED for each object
the maximum contribution from a 500~Myr old SSP allowed by the
$1\sigma$ uncertainties on the $K_s$, $\rm 3.6~\mu m$, and $\rm
4.5~\mu m$ photometric points.  The typical maximum contribution
corresponds to an increase in flux by a factor of 2.5 and translates
into a factor of $\approx 7$ times more stellar mass
\footnote{The additional flux of this old component to the other bands
 always remains within the $1\sigma$ uncertainties of the photometry.
 At observed wavelengths bluer than $K_s$, the light contribution of the
 old component becomes much smaller relative to the strong rest-frame
 UV emission from the young stars.  Compared to the $\rm 3.6$\um\ and
 $\rm 4.5$\um\ data, the $\rm 5.8~\mu m$ and $\rm 8~\mu m$ photometry
 does not provide useful limits because of significantly lower sensitivity 
 and larger uncertainties.}. 

However this factor must not be taken at face value. The assumption of
a 500\myr\ - to maximally-old passively evolving population formed in
an instantaneous burst, made to maximise the mass-to-light ratio, is
more extreme than any detected population.  Any scenario with an old
component that has been forming stars for some non-negligible period
of time, even if short, will reduce this upper limit. And considering
our conservative photometric uncertainties, an upper limit to the mass
contributions for our \zfv\ LBGs will be considerably less than
this. Similar factors have been determined for \zth\ LBGs
\citep[3-5][]{2001ApJ...559..620Papovich,2001ApJ...562...95Shapley},
and therefore such a potential contribution from old stars would be
insufficient to make the mass ranges shown in Section \ref{comp} of
the sample overlap.

\label{lastpage}

\end{document}